\documentclass[a4paper,prl,twocolumn,showpacs,superscriptaddress,nobibnotes]{revtex4-2}
\pdfoutput=1

\usepackage{graphicx}
\usepackage{titletoc}
\usepackage{physics}
\usepackage{enumitem}
\usepackage{bbold}
\usepackage{amsmath, amsthm, amssymb,mathrsfs,bm}

\usepackage{mathtools}

\usepackage[normalem]{ulem}
\usepackage{optidef}
\usepackage{algorithmic}
\usepackage{algorithm}

\usepackage{algorithm}
\usepackage{pifont}
\newcommand{\cmark}{\ding{51}}%
\newcommand{\xmark}{\ding{55}}%

\newcommand{\be}{\begin{equation}}
\newcommand{\ee}{\end{equation}}

\usepackage{dsfont}
\usepackage{vwcol}
\usepackage[colorlinks, citecolor =magenta]{hyperref}
\usepackage[usenames,dvipsnames,table]{xcolor}

\usepackage[acronym]{glossaries}
\makeglossaries
\newacronym{qkd}{QKD}{quantum key distribution}
\newacronym{di}{DI}{device-independent}
\newacronym{pm}{PM}{prepare-and-measure}
\newacronym{sdp}{SDP}{semidefinite programming}
\newacronym{POVM}{POVM}{Positive Operator Valued 
Measure}
\newacronym{usd}{USD}{Unambiguous state discrimination}
\newacronym{qber}{QBER}{Quantum Bit Error Rate}

\usepackage{float}

\usepackage{tikz}
\usetikzlibrary{
  arrows.meta,
  automata
}
\tikzset{
    -Latex,auto,node distance =1 cm and 1 cm,semithick,
    state/.style ={ellipse, draw, minimum width = 0.7 cm},
    point/.style = {circle, draw, inner sep=0.04cm,fill,node contents={}},
    bidirected/.style={Latex-Latex,dashed},
    el/.style = {inner sep=2pt, align=left, sloped}
}

\usepackage{multirow}
\usepackage{diagbox}

\usepackage{wrapfig}

\usepackage[normalem]{ulem}
\newcommand{\stkout}[1]{\ifmmode\text{\sout{\ensuremath{#1}}}\else\sout{#1}\fi}

\usepackage[final]{changes}

\begin{document}

\title{
  Closing the detection loophole in the triangle network with high-dimensional photonic states
}

\author{Tamás Kriváchy}
\email[Contact author: ]{tamas.krivachy@gmail.com}
\affiliation{ICFO - Institut de Ciencies Fotoniques, The Barcelona Institute of Science and Technology, 08860 Castelldefels (Barcelona), Spain}
\affiliation{Atominstitut, Technische Universität Wien, 1020 Vienna, Austria}
\author{Martin Kerschbaumer}
\affiliation{ICFO - Institut de Ciencies Fotoniques, The Barcelona Institute of Science and Technology, 08860 Castelldefels (Barcelona), Spain}
\affiliation{Atominstitut, Technische Universität Wien, 1020 Vienna, Austria}
\affiliation{University of Vienna, Faculty of Physics, Boltzmanngasse 5, 1090 Vienna, Austria}

\begin{abstract}
Bell nonlocality without input settings, e.g. in the triangle network, has been perceived to be particularly fragile, with low robustness to noise in physical implementations. Here we show to the contrary that nonlocality based on N00N states already for $N=2$ has an exceptionally high robustness to photon loss. For the dominant noise factor, single photon loss in the transmission channels, we can certify noise robustness up to 10\% loss, while for a realistic noise model we use neural network-based heuristics to observe $
\sim 50\%$ robustness. Moreover we show that the robustness holds even for imperfect sources based on SPDC sources, where the heralding information of the sources can be used to avoid any global post-processing of the outcomes, such as discarding rounds when photons fail to arrive, and thus demonstrate how the detection loophole in the triangle network can be closed.
\end{abstract}

\maketitle

% \paragraph{Introduction}
The strength of quantum correlations is most profoundly displayed by violations of Bell inequalities, which certify that quantum resource allow for stronger correlations than local causal models~\cite{Bell,brunnerreview}. Naive experimental demonstrations of this effect leave open loopholes~\cite{loopholes_Larsson2014}, allowing for the possibility of local causal theories to explain the observed statistics. Through intense efforts, these loopholes have been closed~\cite{BellDet1_Rowe2001,BellDet2_Ansmann2009,BellDet3_Giustina2013,BellDet4_PhysRevA.90.032107,BellDet5_PhysRevLett.111.130406,BellDet6_PhysRevLett.119.010402,loopholefree1_Hensen2015,loopholefree2_Giustina2015,loopholefree3_Shalm2015}. Besides the closable loopholes, there remain assumptions which must be made; of particular importance is that the choice of measurements that the two parties make are independent of their common past, i.e. of a potentially shared local hidden variable.

\begin{figure}[t]
    \centering
    \includegraphics[width=0.445\linewidth]{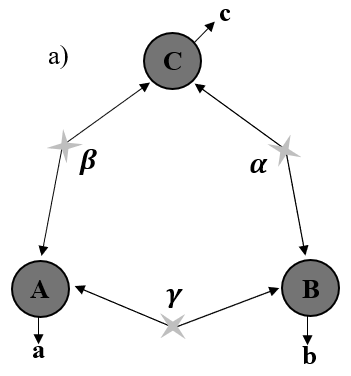}
    \includegraphics[width=0.535\linewidth]{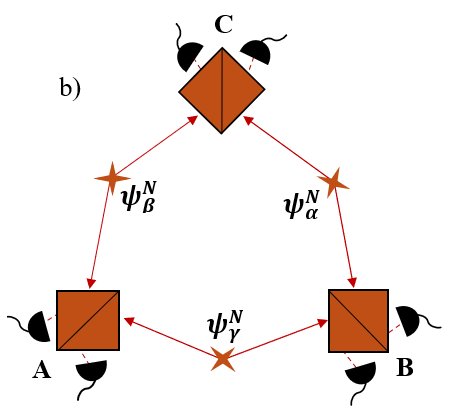}
    \caption{\justifying
    %\justifying (a) The triangle network involves three observers, \( A \), \( B \), and \( C \), with each pair linked by one of three independent bipartite sources \( \{\alpha, \beta, \gamma\} \). These sources distribute physical states to the respective observers. Upon receiving their parts of the states, each observer generates an output through a local process.
    (a) In a triangle network each of the three sources $\alpha,\beta,\gamma$ distributes physical systems to two parties (either quantum or classical). The parties generate a classical output after locally manipulating their received systems.
    %(b) Proposed quantum experiment with imperfect sources: Each source is designed to ideally herald the state \(\ket{\psi}^{N=2}_\xi = \frac{1}{\sqrt{2}} \left( \ket{02} + \ket{20} \right)\) by combining two spontaneous parametric down-conversion (SPDC) sources. 
    (b) In the proposed quantum experiment, each source distributes a N00N state and each party uses a beamsplitter and either photon number resolving or ``click/no-click'' detectors.}
    \label{fig:fig1}
\end{figure}

Fascinatingly, in the network setting one can leave away measurement choices for the parties, at the cost of assuming that the local hidden variables shared between certain parties are independent, see e.g. the structure of the triangle network in Fig.~\ref{fig:fig1}(a)~\cite{Fritz2012,Renou2019,tavakoli_bell_2022}.
As in the early days of standard Bell tests, current network nonlocality experiments suffer from certain loopholes which should be closed, of which a critical one is the detection loophole. Next to low detection efficiencies (or high losses in the channels), nonlocality wanes~\cite{krivachy_neural_2020,abiuso_single-photon_2022}, forcing the use of global post-processing of the outcomes in order to keep only those rounds in which all photons arrived, in the spirit of the fair sampling assumption~\cite{polino_experimental_2023,wang2024experimentalgenuinequantumnonlocality}. Global post-processing such as this is forbidden, however, as by discarding certain rounds based on global information of all parties' outcomes in a network, one can effectively introduce correlations between the parties and recreate any distribution, i.e. one opens a loophole.

Previous proposals relying on single-photon sources for nonlocality in the triangle had $\sim 5\%$ noise robustness~\cite{abiuso_single-photon_2022} (0.4\% proven~\cite{boreiri_noise-robust_2023}), where the number indicates the probability of losing a photon in a single transmission mode from a source to a party. These figures are well below the coupling loss of $\sim20\%$ that is present in experiments which couple optical modes into fibers for the purpose of using single-photon detectors.

Inspired by the exceptional noise robustness of high-dimensional quantum states in quantum information tasks~\cite{noiseVertesi2010_closing,noise_exp_PhysRevLett.129.060402,noisePhysRevX.9.041042,noisePhysRevX.12.041023,noise_qudit_summary,noiseDaLio2021,noiseZhu2021}, we examine the a 2-photon source state and passive optical elements (see Fig.~\ref{fig:fig1}(b)), which displays $\sim 50\%$ robustness according to numerical heuristics. We reinforce these results by generalizing previous nonlocality proofs, establishing a framework for proving nonlocality beyond the standard ``token counting'' technique. Our technique is more adapted to token (photon) loss, allowing us to study the dominant noise factor, single photon loss in the transmission channels, and certify that our setup's nonlocality is robust up to 10.3\% single-photon loss, and up to 0.14\% full photon loss noise. The order of magnitude jump in both heuristically observed noise robustness (from $\sim$5\% to $\sim$50\%) and in certified robustness against the dominant noise source, single photon loss (from $0.4\%$ to $10.3\%$), show that using high dimensional (qudit) sources in network nonlocality is indeed promising for moving from theory to the laboratory.

Moreover, we consider imperfection of the sources on a specific spontaneous parametric down-conversion (SPDC) source-based model. We address both higher-order effects and show how\added{, by} using the heralding information of the sources\added{,} the vacuum component can be disregarded. Together with the robustness against photon loss, these completely eliminate the need for the parties to conduct global post-processing of the rounds due to photon non-arrival events, thus closing the detection loophole.

%In principle, the detection loophole can thus be closed if one has a perfect N00N state. However, imperfections may arise that affect nonlocality, depending on the implementation. As a proof of concept, we show how nonlocality from this proposal is robust to imperfections in a source generated via spontaneous parametric down-conversion (SPDC) sources. First we demonstrate that higher-order effects in the source generation do not significantly affect the noise robustness. Then we show how using the heralding information of the sources the vacuum component can be disregarded. This, together with the robustness against photon loss, completely eliminates the need for the parties to conduct global post-processing of the rounds due to photon non-arrival events, thus closing the detection loophole.

% In the manuscript we first introduce the scenario, the proposed quantum strategy and the generalization of the previous nonlocality proofs. Next, we study noise robustness under photon loss, but with perfect sources. Then we show that heralded source failures do not influence nonlocality of the distribution, and discuss the influence of SPDC imperfections. Finally, we conclude with a discussion and open questions.

% \paragraph{Nonlocality under photon loss}
\paragraph{Nonlocality under photon loss--}
In general, a triangle with classical sources leads to a distribution of the form
\begin{align}
    p(a,b,c) = \int d\widetilde\alpha d\widetilde\beta d\widetilde\gamma & p_A (a|\widetilde\beta,\widetilde\gamma)  p_B (b|\widetilde\gamma, \widetilde\alpha) p_C (c|\widetilde\alpha,\widetilde\beta) \nonumber \\  & \cdot p_{\alpha}(\widetilde\alpha) p_{\beta}(\widetilde\beta) p_{\gamma}(\widetilde\gamma),
    \label{eq: triangle_LHV}
\end{align}
where $p_{\xi}$ is the probability distribution of the classical random variable distributed from source $\xi \in\{\alpha,\beta,\gamma\}$ and the conditional probability $p_X(x|\widetilde\xi_1,\widetilde\xi_2)$ is the response function of party $X\in\{A,B,C\}$.

In contrast, if the sources \deleted{can }distribute quantum states then the parties can sample from distributions of the form
\begin{equation}
    \label{eq: triangle_quantum}
    p(a,b,c) = \text{Tr}(\rho_{\alpha} \rho_{\beta} \rho_{\gamma} M_A^a M_B^b M_C^c),
\end{equation}
where $\rho_{\xi}$ is the state distributed by source $\xi$ and $M_X^x \geq 0$ is the POVM element corresponding to output $x$ of party $X$, such that $\sum_x M_X^x = \mathbb{I}$. Note that we left the tensor products away for brevity (e.g. $M_A^a \otimes M_B^b \otimes M_C^c$ ) and attention must be paid to match the proper Hilbert spaces (e.g. $A \equiv (A_1, A_2); \beta \equiv (A_2, C_1)$).

Whereas previous examples of nonlocality in the triangle rely on the distribution of qubits from the sources, here we propose that each source $\xi$ distributes a pair of qudits in a so-called NOON state
\begin{equation}
    \ket{\psi}_{\xi}^N = \frac{1}{\sqrt{2}} (\ket{0N} + \ket{N0}),
\end{equation}
with $N\geq2$, $\rho_{\xi}=\ket{\psi}_{\xi}^N \bra{\psi}_{\xi}^N$, and where the numbers represent photon numbers (Fock basis), i.e. each source creates a superposition of $N$ photons sent to the right and $0$ to the left and $N$ photons to the left and $0$ to the right. The parties then locally combine their two incoming modes on \deleted{a }beam splitters, characterized by a unitary $U$, which depends on a transmissivity coefficient $t$ and a phase shift $\varphi$, and measure the two outgoing modes, resulting in measurement operators
\begin{align}
M^{i_1,i_2}_X = U^\dagger D^{i_1}_{X_1} \otimes D^{i_2}_{X_2} U.
\end{align}
The parties may use either photon number resolving detectors ($D^{i} = \ket{i}\bra{i}$), or ``click/no click'' detectors, which fire if any number of photons arrive and do not fire otherwise ($D^{0} = \ket{0}\bra{0}$, $D^1 = \mathbb{I}-D^0$). In the following we consider all parties to use the same parameters $t,\varphi$ for their measurements and we focus on the $N=2$ case. Then, using photon number resolving detectors one obtains
\begin{align}
    p(0,2_i,4) =& \frac{1}{8} |u_i|^2,\;\; p(0,4,2_i) = \frac{1}{8} |v_i|^2,\\
    p(2_i,2_j,2_k) =& \frac{1}{8} |u_i u_j u_k + v_i v_j v_k|^2,\\
    u_{-1} =  e^{-i2\varphi} t,&\;\; u_1 = e^{-i2\varphi}(1-t),\\
    u_0 = -e^{-i2\varphi}&\sqrt{2t(1-t)},\;\; v_i=|u_{-i}|,
\end{align}
where the outcomes $2_{-1},2_0,2_1$ represent that $(2,0),(1,1)$ or $(0,2)$ photons arrive to the (left detector, right detector), respectively. Note that $0\equiv (0,0)$ and for the 4 outcome we do not care where the photons arrived, as it does not influence the nonlocality conditions we work with. See the Appendix C for a detailed derivation also for tilted N00N states and for click/no-click detectors.

\begin{table*}[t]
    \centering
    \begin{tabular}{|*{7}{c|}}
        \hline
        & \multicolumn{2}{|c|}{Dephasing (or no) noise} & \multicolumn{2}{|c|}{$N-1$ photon loss} & \multicolumn{2}{|c|}{Full photon loss}\\
         & PNRD & C/no-C & PNRD & C/no-C &PNRD & C/no-C
         \\\hline %769
        TC~\cite{Renou2019,renou_tokencounting} & \cmark & \xmark & \xmark & \xmark  & \xmark & \xmark%\\
        %$N=2$ NL for & $dephnoise \leq 66.0\%$ &  &  &   &  & & & \\
        %  & ($t=0.99$) &  &  &   &  & & &
         \\\hline
         nPTC~\cite{boreiri_noise-robust_2023} & \cmark & \xmark & \cmark (only tilted) & \xmark & \cmark (only tilted) & \xmark
         %\\
         %$N=2$ NL for & $dephnoise \leq 66.0\%$ &  & $\eta>...$ &   & $Q<...$  & & $\eta>...$ & 
         \\\hline
         ours & \cmark $o^*\in\{0,4\}$ & \cmark $o^*=0$ & \cmark $o^*=0$ & \cmark $o^*=0$ & \cmark $o^*=4$ & \xmark \\
         $N=2$ NL for & d$\leq 66\%$ & d$\leq 66\%$ & $\eta\geq 89.7\%$ & $\eta\geq 93.3\%$ & $\eta\geq 99.86\%$& \\
          & ($t=0.99$) & ($t=0.99$) & ($t=0.97$) & ($t=0.94$)  & ($t=0.82$) & 
         \\\hline
    \end{tabular}
    \caption{Summary of techniques for proving nonlocality, and for which distribution (photon number resolving detector (PNRD) or click/no-click (C/no-C) detector) and for which noise models they can be applied (\cmark) or not (\xmark) (Token Counting (TC), noisy Parity Token Counting (nPTC) and ours; numerical values given for $\varphi=\frac{\pi}{2}$).
    %, $Q$ is the probability of more than one photon pair emitted from the SPDC source.
    }
    \label{table:LPsummary}
\end{table*}
% Table 1: for using TC dephasing is limited at 66\% for N=2. However, when using our LP it is larger (note we use 0,(20),(11),(02),4 cg. Why? For N=1 seems to not make a difference. OK probably it is related to the extreme sensitivity of dephasing noise threshold to numeric values. So we keep 66\% in the table.

The most relevant noise sources 
%in practical setups based on discrete variable photonics
are photon loss and source failures, opening loopholes in experiments, as these issues can only be remedied via global postprocessing, e.g. by globally rejecting all events where an incorrect number of photons arrived. Note that global rejection can introduce incredibly strong correlations, e.g. one can achieve GHZ-like correlations from local binary-outcome coin tosses by discarding all events besides (0,0,0) and (1,1,1) outcomes, which is otherwise impossible~\cite{Renou_quantumFinner_2019}. Thus, it is best avoided in network nonlocality analysis.
%In the current manuscript we provide a technique for circumventing closing the former loophole and give strong evidence that the current experimental proposal for $N=2$ is robust against photon loss, to a level that is achievable with current experiments.
%We consider a noise model that captures the previous two aspects.

First, let \deleted{a}\added{u}s assume a perfect source and study photon loss in the channel between the sources and the parties via the \textit{full photon loss} noise model for $N=2$
% \begin{align}
%     &E_0 = \ket{0}\bra{0} + \sqrt{\eta}\ket{1}\bra{1} + \eta \ket{2}\bra{2},\\
%     &E_1 = \sqrt{1-\eta} \ket{0}\bra{1} + \sqrt{2 \eta (1-\eta)}\ket{1}\bra{2},\\
%     &E_2 = (1-\eta) \ket{0}\bra{2},\\
%     &M^{(i_1,i_2)}_\eta = \sum_{j,j'} E^\dagger_j\otimes E^\dagger_{j'} U^\dagger D_{i_1} \otimes D_{i_2} U E_j\otimes E_{j'},
% \end{align}
\begin{align}
    &E_0 = \ket{0}\bra{0} + \eta \ket{2}\bra{2},\\
    &E_1 = \sqrt{2 \eta (1-\eta)}\ket{1}\bra{2},\\
    &E_2 = (1-\eta) \ket{0}\bra{2},\\
    M^{(i_1,i_2)}_\eta =& \sum_{j,j'} E^\dagger_j\otimes E^\dagger_{j'} M^{i_1,i_2} E_j\otimes E_{j'},
\end{align}
where $1-\eta$ is the probability of a single photon being lost, $i_1,i_2\in\{0,\dots,2N\}$ for photon number resolving detectors and $i_1,i_2\in\{0, 1\}$ for click/no-click detectors, and where we have omitted the $\ket{\cdot}\bra{1}$ terms as $\ket{1}$ does not appear in the source state. We consider another noise model which allows us to capture the most dominant factor of noise, \textit{single photon loss},
\begin{align}
    &E_0 = \ket{0}\bra{0} + \eta \ket{2}\bra{2},\\
    &E_1 = \sqrt{1-\eta^2}\ket{1}\bra{2},
\end{align}
where each mode can lose maximally one photon, with a probability of successful transmittion of a single photon of $\eta$.
% Note there is a t <-> 1-t swap in this with respect to my code

Finally, \deleted{ though it is less relevant experimentally,}we consider dephasing noise, where each source is replaced by 
\begin{align}
    (1-d) \ket{\psi}^N\bra{\psi}^N + \frac{d}{2} (\ket{0N}\bra{0N}+\ket{N0}\bra{N0}).
\end{align}
Similarly to the $N=1$ case~\cite{boreiri_noise-robust_2023}, the robustness to dephasing noise is surprisingly high.

\noindent\textbf{Result 1.} The above family of distributions for $N=2$ is nonlocal in the triangle for $\varphi=\frac{\pi}{2}, 0.761\leq t < 1$ for both photon number resolving detectors and click/no-click detectors, and are robust to the $N-1$ photon lost noise model and full photon loss noise model as specified in Table~\ref{table:LPsummary}.

\added{Further regions of nonlocality are provided in Appendix B. }The proof is akin to the Token Counting (TC) proof developed in~\cite{Renou2019,renou_tokencounting, renou_genericNWs}. However, strict TC does not hold when one uses ``click/no-click'' detectors, or if some photons are lost. In Ref.~\cite{abiuso_single-photon_2022} a proof was developed to accommodate ``click/no-click'' detectors for $N=1$ without noise. Here we provide a generalization which allows the accommodation of photon loss noise models as well (see Table~\ref{table:LPsummary}). Note that another generalization of TC exists which can accommodate noise~\cite{boreiri_noise-robust_2023}, however our technique preserves the structure of the problem much better for the distributions under examination\added{, since the central structure established in Lemma 1 remains invariant even under certain variations in the number of tokens (or photon counts), whereas ~\cite{boreiri_noise-robust_2023} treats all such variations as noise}. A full proof of Result 1 is given in Appendix B; here we only highlight the central ingredients in the generalization.

\noindent\textbf{Lemma 1.} Let $p(a,b,c)$ be a triangle-local distribution for which there exists some outcome label $o^*$ s.t.\added{ $o^*$ never occurs at two parties simultaneously, i.e.}
\begin{align}
    p(o^*,o^*,c)=0 \;\; \forall c,\nonumber\\
    p(a, o^*,o^*)=0\;\; \forall a,\label{eq: property of Lemma 1}\\
    p(o^*,b,o^*)=0\;\; \forall b.\nonumber
\end{align}

% \begin{align}
%     p(a=b=o^*)=0,\nonumber\\
%     p(b=c=o^*)=0,\nonumber\\
%     p(c=a=o^*)=0.\nonumber
% \end{align}
Then \added{each latent variable essentially consists of a ``Left'' and a ``Right'' part}, i.e. for the domain $S_{\xi}$ of each source $\xi$'s latent variable \added{($\xi\in\{\alpha,\beta,\gamma\}$)} it must hold that $S_{\xi} = S_{\xi}^L \cup S_{\xi}^R$, such that $S_{\xi}^L \cap S_{\xi}^R = 0$\added{,} and \added{the outcome $o^*$ can only occur if a party receives a symbol from $S^L$ from its left source and a symbol from $S^R$ from its right source, i.e. $p_X(o^*|\widetilde \xi_1,\widetilde \xi_2) =0$ if $\widetilde\xi_1 \notin S_{\xi_1}^L$ or $\widetilde\xi_2 \notin S_{\xi_2}^R$.}\deleted{$p_X(o^*|\widetilde \xi_1,\widetilde \xi_2) >0$ if $\widetilde\xi_1 \in S_{\xi_1}^L$ and $\widetilde\xi_2 \in S_{\xi_2}^R$.} Moreover\added{,} if $p(o^*):=p(a=o^*)=p(b=o^*)=p(c=o^*)$ then
% \begin{align}
%     |S_{\xi}^L| \in [\frac{1}{2} - \frac{1}{2}\sqrt{1-4p(o^*)},\frac{1}{2} + \frac{1}{2} \sqrt{1-4p(o^*)}]
% \end{align}
\begin{align}
    \left|p\left(\tilde\xi \in S_{\xi}^{(L,R)}\right) - \frac{1}{2}\right| \leq \frac{1}{2}\sqrt{1-4p(o^*)}
    \label{eq: ineq. in Lemma 1}
\end{align}
and if $p(o^*)=\frac{1}{4}$ then $p_X(o^*|\tilde\xi_1,\tilde\xi_2) >0$ if and only if $\widetilde\xi_1 \in S_{\xi_1}^L$ and $\widetilde\xi_2 \in S_{\xi_2}^R$.

%must have a local model where each source's symbols are divided in two disjoint sets: those that allow for $o^*$ to occur for the party on the right, and those that allow for $o^*$ to occur for the party on the left. The parties output $o^*$ if and only if they receive symbols permitting them to output $o^*$ from both sources.}

% More mathematically:
% $\xi \in S_{\xi} = \{1, 2, \dots M\}$
\noindent \textit{Proof sketch of Lemma 1.} Lemma 1 follows from a simple geometric argument, portrayed in Appendix B. The main idea is that if Alice outputs $o^*$ for some values $\tilde\beta,\tilde\gamma$ of her hidden variables, then Bob and Charlie are forbidden from outputting $o^*$ when seeing any of those variables. Then, by circularly applying this argument one sees that as $p(o^*)\rightarrow1/4$, there is less and less flexibility in the local model, until finally complete rigidity kicks in at $1/4$. 

\noindent \textit{Proof sketch of Result 1.} For the noiseless distributions under examination\added{,} Lemma 1 can be used with $o^*=0$ (note that for the photon number resolving case it simultaneously applies for $o^*=0$ and $o^*=2N$), with $p(o^*)=1/4$.
%As such, the local models have a strict structure, with six possible combinations which result in outcomes with $o^*$ in them: $(\alpha,\beta,\gamma)$ respectively coming from $(S_{\alpha}^L, S_{\beta}^L, S_{\gamma}^R)$, 
% $(S_{\alpha}^L, S_{\beta}^R, S_{\gamma}^L)$, $
% (S_{\alpha}^L, S_{\beta}^R, S_{\gamma}^R)$, $
% (S_{\alpha}^R, S_{\beta}^L, S_{\gamma}^L)$, $
% (S_{\alpha}^R, S_{\beta}^L, S_{\gamma}^R)$, $
% (S_{\alpha}^R, S_{\beta}^R, S_{\gamma}^L)$. 
% The remaining two options, $(S_{\alpha}^L, S_{\beta}^L, S_{\gamma}^L)$, $(S_{\alpha}^R, S_{\beta}^R, S_{\gamma}^R)$, by Lemma 1, can not result in an $o^*$ outcome appearing, thus all other outcomes must come from these subspaces. 
As such, the local models have a strict structure in which the result $o^*$ is strictly forbidden from appearing when all hidden variables come from the same types of sub-domains, i.e. $(\widetilde\alpha,\widetilde\beta,\widetilde\gamma)$ belongs to $(S_{\alpha}^L, S_{\beta}^L, S_{\gamma}^L)$ or $(S_{\alpha}^R, S_{\beta}^R, S_{\gamma}^R)$. All (non-$o^*$, non-$o^*$,non-$o^*$) events must be from these sub-domains.
%Note that for the strict TC proofs these other outcomes are strictly those with $N$ photons arriving at each party, however, as was generalized in ... and here, these outcomes can have a different number of photons/tokens, allowing the use of click/no-click detectors and the accommodation of loss of certain photons (tokens), as we discuss in Sec.....
Finally, using the principles that, for example, party A's outcome can not be influenced by $\alpha$, one can set linear constraints that must be satisfied for how these non-$o^*$ events are distributed among the two subdomains. Such constraints must be satisfied by any local model, and incompatibility of the constraints can be verified by a linear program (LP) in the regime stated above. $\square$ %\tamas{clockwise/counter-clockwise?}

% \paragraph{Robustness under photon loss}
% Currently the largest obstacle of conducting a loophole-free test of nonlocality in the triangle network is noise. Hence, noise robustness analysis is a crucial part of examining experimental proposals such as the current one. Most analytic techniques can give only minor noise robustness guarantees under physical noise models, hence numerical heuristics have become a popular alternative to gauge the true noise robustness of distributions, and seem to provide consistent and reliable estimates.

 Note that in both photon loss models $p(4)<\frac{1}{4}$ for $\eta<1$, which weakens the use of Lemma 1. For full photon loss $o^*=0$ may not be used with Lemma 1, as $p(a=0,b=0)>0$ for $\eta <1$.  However, in the single photon loss model, the 0 events remain unaffected by the noise model, i.e. Lemma 1 may still be used with $o^*=0$ and $p(0)=\frac{1}{4}$. In general the technique to approximate the full $N$-photon loss model with $N-1$-photon loss leaves the 0 outcomes untouched and allows for a strong use of Lemma 1. % We summarize the applicability of proof techniques in Table~\ref{table:LPsummary}.
 \added{Furthermore, one could easily incorporate dark count rates of detectors, as these do not affect the condition in Lemma 1 for $o^*=0$ as they only increase the number of counts and thus don't generate $(a=0,b=0,c)$-type events. However, for our current proof of principle experiment with heralded sources (see below), we expect dark count contributions to be minimal, since coincidences must only be studied in very small time windows, on the order of the jitter of the detectors.}

\begin{figure}[t]
    \centering
    \includegraphics[width=0.82\linewidth]{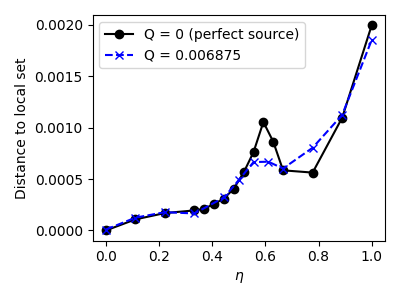}
    \caption{\justifying Distance to the local set according to LHV-Net for different values of channel transmittivity $\eta$ and source impurity  $Q$, for the full photon loss model with click/no-click detectors, for $t=0.75, \varphi=\frac{\pi}{2}.$} 
    \label{fig:LHVNet}
\end{figure}

% The above values are lower bounds to the noise robustness, so i
Moving beyond the results in Table~\ref{table:LPsummary}, in order to gauge the true noise robustness of the distributions, we use a neural network-based representation of local models of the form (\ref{eq: triangle_LHV})~\cite{krivachy_neural_2020}. \added{The neural network (}LHV-Net\added{)} can only generate local distributions, and is asked to minimize the Euclidean distance to a given target distribution. For the results in Fig.~\ref{fig:LHVNet} we retrain the network for a fixed $t=0.75, \varphi=\frac{\pi}{2}$, but for different values of $\eta$ in order to see where a transition appears between local and nonlocal distribution, where for the former LHV-Net can approach an almost 0 distance and for the latter the distances are larger~\cite{Note1}. Starting from the noiseless $\eta=1$, one sees that the Euclidean distance decreases as $\eta$ reduces. Interestingly, a resurgence appears around $\eta\approx 0.6$, where the distance is larger than before, and then decreases again as noise becomes dominant. Identifying the precise noise tolerance threshold for nonlocality is difficult from such heuristics, however the existence of a robust resurgent peak at $\eta\approx 0.6$ shows us that the true noise robustness of the experiment almost certainly exceeds $40\%$, well above the photon loss rate achievable in today's tabletop experiments. Details of the LHV-Net implementation are given in Appendix E.
%\footnote{Note that even though the LP proof doesn't provide infeasibility for $t=0.75$, we expect it to exist based on results in \cite{pozas_continuous_families}. We chose this $t$ value since LHV-Net found this region to be farthest from the local set.}

\paragraph{Robustness under source imperfection--}
The previous paragraphs address nonlocality under any implementation of the $\ket{\psi}^N$ state. Here we consider its implementation with SPDC sources, which is often used in quantum optics experiments, through which a state $\ket{\psi} \propto \sum_{k=0}^\infty q^k\ket{kk}$ can be created. As shown in detail in Appendix D, one of each of the outgoing modes of the SPDC sources is measured in order to herald the creation of a photon in the other mode. Then ideally, two heralded single photons combine on a beam splitter, where through the Hong--Ou--Mandel effect, a state $\ket{\psi}^{N=2}$ is created.

In reality, conditioned on detecting a single photon, the state before the Hong--Ou--Mandel combination can be approximated by $(1-Q)\ket{1}\bra{1} + Q\ket{2}\bra{2} + \mathcal{O}(Q^2)$. Combining two of these states on a beam splitter leads to an imperfect $\ket{\psi}^{N=2}$, where for the calculations we kept only terms up to order $Q^2$. Following Ref.~\cite{abiuso_single-photon_2022}, we considered $Q=0.006785$ to be an approximately realistic value, which we utilized in Fig.~\ref{fig:LHVNet}. The previous robustness to photon loss is barely influenced by the presence of such higher order terms in the source.

Finally, we show how heralded generation of entanglement can be used to circumvent the need for global post-processing, exemplified through the previous SPDC setup, modified according to the following. %As illustrated in Fig \ref{fig:fig1}, one of each of the outgoing modes of the SPDC sources is measured in order to herald the creation of a photon in the other mode. Then ideally, two heralded single photons combine on a beam splitter, where through the Hong--Ou--Mandel effect, a state $\frac{1}{\sqrt{2}}(\ket{02}+\ket{20})$ is created.
%\footnote{Note that we do not consider higher-order photon generation events; imperfections from these can in principle be suppressed by the use of photon number resolving heralding detectors.}
In each round, each source has a shutter blocking the path of the photons to the parties, which the source removes if and only if both SPDC processes were heralded as successful. In the case of successful source generation, the source sends a bit `0' to both parties, otherwise, if any of the heralding processes fail, then it sends a failure bit `1' to the parties. One can immediately see that in such a setup, token counting is preserved if one considers the bit values as tokens as well, since the source emits either 2 photons and 0's in the bit registers or 0 photons and two classical `1' values. Thus, for our distribution with photon number resolving detectors the same nonlocality proof holds as before. However, the scheme is more powerful, since the failure bit coming from the sources strictly constrains any local hidden variable model.

\noindent\textbf{Lemma 2} If each source $\xi$ can certify its failure, then by sending the additional failure bit $f_{\xi}$ to the parties they are connected to, the probability distribution $p\left((a,f_{\beta}, f_{\gamma}), (b,f_{\gamma}, f_{\alpha}), (c, f_{\alpha}, f_{\beta})\right)$ is local if and only if $p(a,b,c)$ is local.

We give a simple geometric proof in Appendix D. The consequence of this Lemma is that our previous nonlocality results (even those with photon loss noise) hold even if the sources fail sometimes in a certifiable fashion. Thus, global postprocessing in order to disregard failure of SPDC sources is not necessary in the proposed experimental setup.

While finishing the manuscript, we found that similar considerations regarding failures of heralded sources in general have very recently been addressed in Ref~\cite{boreiri2025bellnonlocalityquantumnetworks}.

\paragraph{Discussion--}
As interest in experimental demonstrations of nonlocality in networks without inputs grows~\cite{polino_experimental_2023,wang2024experimentalgenuinequantumnonlocality,ivan_experiment_meskine2024experimentalquantumtrianglenetwork}, it has become increasingly timely to address possible loopholes in such implementations. We have proposed a distribution which shows that nonlocality in the triangle network is not as fragile as it seems, and the detection loophole can be closed, through the assistance of high-dimensional quantum states. Naturally, other loopholes must still be addressed, e.g. those related to shared reference frames~\cite{sharedRefFrames_PhysRevA.95.062315} or memory attacks~\cite{weilenmann2024memory}.

We have studied this scenario with complementary tools, getting both certificates with the LPs, and a more realistic picture of the true noise robustness with LHV-Net.
It has been noticed in previous works as well, that often the certified noise robustness based on an LP is an order of magnitude smaller than real one, naturally since these LP's capture only some aspects of the full local constraints. We observe this here as well. In the spirit of Ref.~\cite{pozas_continuous_families}, we could possibly improve these results by using the inflation technique on the feasible solutions of the LP's. However, this is challenging due to the increased number of outcomes (for $N=2$, naively 4 outcomes per party are needed for the noise models vs. the 2 that were required in~\cite{pozas_continuous_families}).

Note that the results from LHV-Net exhibit a clear example that the Euclidean distance to the local set is not a physically meaningful measure of the strength of nonlocality, as it can increase under the addition of physical noise. Though in a different context, such nonlocal ``bumps'' can be observed elsewhere as well~\cite{baumer2024exploring}.
For the current distribution, the resurgent bump appearing in the noise plot could be related to the fact that the $N=1$ distribution is so robust to dephasing~\cite{boreiri_noise-robust_2023}. In particular, consider the case when the $N=2$ source undergoes a photon loss on one of the sides. Then the state decays as $
\frac{1}{\sqrt{2}}(\ket{02}+\ket{20}) \rightarrow \frac{1}{2} \left(\ket{01}\bra{01} + \ket{10}\bra{10}\right)$, arriving at a dephased $N=1$ source state. This, however, remains just an intuition and profoundly understanding the strong robustness remains an open question\added{ (we discuss additional considerations about the resurgent peak in Appendix E)}. Further\deleted{more} interesting theoretic questions remain for this setup, such as whether the resurgent nonlocal bumps appear for other $N$ values, and for what values of noise do properties like genuine network nonlocality, full network nonlocality or topological robustness appear~\cite{supic_genuine_NNL2022, pozas_full_network_nonlocality_2022, sekatski_boreiri_partial_2023, boreiri_krivachy_sekatski_topRobNL_2024}.

Finally, the finding \added{of }such robustness to photon loss would be of interest both for implementations of other distributions such as the Elegant distribution, strongly believed to be nonlocal~\cite{gisin_published_ejm,krivachy_neural_2020,baumer2024exploring,wang2024experimentalgenuinequantumnonlocality,numerical_elegant_daSilva_Parisio_2023}, or for other realizations of token-counting (and related) distributions. In particular, with the rapid advance of quantum technologies, e.g. on demand photon emission, it is an interesting question whether analogous effects can be observed with such other technologies.

\paragraph{Acknowledgments--} We are grateful to Paolo Abiuso, Alejandro Pozas-Kerstjens, Antonio Acín and Marcus Huber for useful inputs on the project. We acknowledge financial support from the Government of Spain (Severo Ochoa CEX2019-000910-S and FUNQIP), Fundació Cellex, Fundació Mir-Puig, Generalitat de Catalunya (CERCA program). T.K. additionally acknowledges funding from the Swiss National Science Foundation (project 214458) and from the Austrian Federal Ministry of Education via the Austrian Research Promotion Agency–FFG (flagship project FO999897481, funded by EU program NextGenerationEU). M.K. additionally acknowledges funding from the Erasmus+ Programme of the European Union through an Erasmus+ traineeship grant, as well as from the Gesellschaft für Forschungsförderung Niederösterreich m.b.H.

\paragraph{Data availability--} All data can be generated based on the manuscript, particularly with the assistance of sample code openly available at~\cite{code} for setting up and executing the LP, as well as for running LHV-Net.
\nocite{minimal_NL_boreiri_PhysRevA.107.062413}

% \bibliographystyle{unsrt}
% \bibliography{main.bib}

\newpage
\onecolumngrid
\appendix
\newpage
\section{A. Cube representation of LHV models}
\added{In the classical setting all randomness in the local statistical response functions can w.l.o.g. be encoded in the sources, thus the local response functions can be considered deterministic (any randomness used by a given party could be generated and passed on by one of the sources). Furthermore, w.l.o.g. the hidden variables \(\{\widetilde{\alpha},\widetilde{\beta},\widetilde{\gamma}\}\) can be taken to be uniformly distributed over the interval \([0,1]\), as any distribution can be reached via a local postprocessing of such randomness by the parties e.g. via the inverse transform sampling method. As a result finding a local model is equivalent to finding the deterministic responses of each of the parties, i.e. $p_A(a|\widetilde\beta,\widetilde\gamma), p_B(b|\widetilde\gamma,\widetilde\alpha), p_C(c|\widetilde\alpha,\widetilde\beta)$. This further allows us to view the local model in a geometric cube picture, which we introduce here.}

\deleted{In the classical setting, the hidden variables \(\{\widetilde{\alpha},\widetilde{\beta},\widetilde{\gamma}\}\) can w.l.o.g. be taken to be uniformly distributed over the interval \([0,1]\). Moreover, all randomness in the local statistical response functions can be encoded sources, thus again without loss of generality, the local response functions can be considered deterministic.}

\begin{figure}[b]
    \centering
    \includegraphics[width=\linewidth]{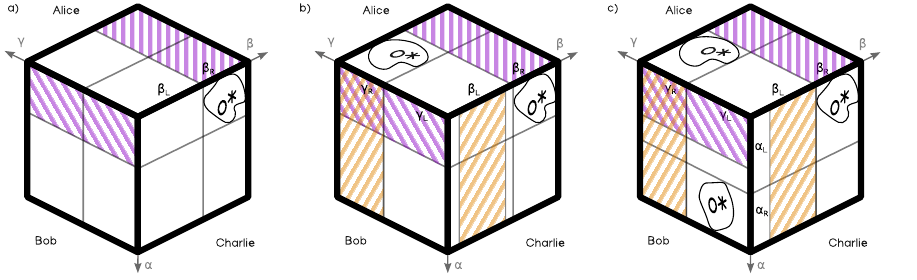}
    \caption{Cube representation of any LHV model which could possibly reproduce p(a,b,c). (a) First, we rearrange $\beta$ to collect all \(o^*\) outputs from Charlie in one region, defining $\beta_R := p(S_\beta^R)$. Consequently, \deleted{Bob is prohibited from outputting \(o^*\) within the region \(\alpha_1\), and Alice is restricted from doing so within the region \(\beta_2\), both of which are highlighted in purple} Alice and Bob are prohibited from outputting \(o^*\) within the striped purple regions. b) Next, we repeat the same procedure for $\gamma$ and Alice's $o^*$, defining $\gamma_R$\added{$:=p(S_\gamma^R)$}. For Bob and Charlie, regions where an \(o^*\) output is not possible are identified and marked in yellow. (c) Finally, for Bob, we rearrange $\alpha$ to get all $o^*$ outcomes on the bottom and $\alpha_R:=p(S_\alpha^R)$, leading to the inequalities \(\alpha_L \beta_R \geq p(b = o^*), \beta_L \gamma_R \geq p(b = o^*), \gamma_L \alpha_R \geq p(b = o^*)\), which lead to the inequality in Lemma 1. }
    \label{fig:cube_app_lemma1}
\end{figure}

%Each hidden variable can be visualized as a unit-length line segment.
Each response function depends on two hidden variables which can be represented by a two-dimensional unit square where the axes are the corresponding hidden variables. Extending this construction to all response functions in the network, the full set of classical strategies can be represented within a normalized three-dimensional cube (see e.g. Fig.~\ref{fig:cube_app_lemma1}). In this representation, putting labels on the three sides corresponds the parties having those outputs for the corresponding values of the hidden variables. The volume in the cube which is behind an $(i,j,k)$ labeling corresponds to $p(a=i,b=j,c=k)$. See e.g.~\cite{Renou_quantumFinner_2019,baumer2024exploring,boreiri_noise-robust_2023} for examples of the use of the cube model.

Note that due to the integrals, the order of the hidden variables' values doesn't matter, namely one can always swap $\widetilde\alpha_1$ with $\widetilde\alpha_2$, as long as the response functions are adjusted accordingly. This corresponds to rearranging slices of the cube and does not affect the generated $p$.

\section{B. Proof of nonlocality}\label{app:proof_result1}
We denote by $\chi$ the set of all outcomes excluding $o^*$, and $\chi_i$ as an element of this set. In the following, we use $p(a=\chi)$ and $p(a\in\chi)$ equivalently. As such, something like $p(\chi_i,\chi,\chi)$ actually means $p(a=\chi_i,b\in\chi,c\in\chi)$.
\subsection{Proof of Lemma 1}

\added{\noindent\textbf{Lemma 1. (restated)} Let $p(a,b,c)$ be a triangle-local distribution for which there exists some outcome label $o^*$ s.t. $o^*$ never occurs at two parties simultaneously, i.e.}
\begin{align}
    p(o^*,o^*,c)=0 \;\; \forall c,\nonumber\\
    p(a, o^*,o^*)=0\;\; \forall a,\label{eq:app:lemma1_conditions}\\
    p(o^*,b,o^*)=0\;\; \forall b.\nonumber
\end{align}

% \begin{align}
%     p(a=b=o^*)=0,\nonumber\\
%     p(b=c=o^*)=0,\nonumber\\
%     p(c=a=o^*)=0.\nonumber
% \end{align}
\added{Then each latent variable essentially consists of a ``Left'' and a ``Right'' part, i.e. for the domain $S_{\xi}$ of each source $\xi$'s latent variable ($\xi\in\{\alpha,\beta,\gamma\}$) it must hold that $S_{\xi} = S_{\xi}^L \cup S_{\xi}^R$, such that $S_{\xi}^L \cap S_{\xi}^R = 0$, and the outcome $o^*$ can only occur if a party receives a symbol from $S^L$ from its left source and a symbol from $S^R$ from its right source, i.e. $p_X(o^*|\widetilde \xi_1,\widetilde \xi_2) =0$ if $\widetilde\xi_1 \notin S_{\xi_1}^L$ or $\widetilde\xi_2 \notin S_{\xi_2}^R$. Moreover, if $p(o^*):=p(a=o^*)=p(b=o^*)=p(c=o^*)$ then}
% \begin{align}
%     |S_{\xi}^L| \in [\frac{1}{2} - \frac{1}{2}\sqrt{1-4p(o^*)},\frac{1}{2} + \frac{1}{2} \sqrt{1-4p(o^*)}]
% \end{align}
\begin{align}
    \left|p\left(\tilde\xi \in S_{\xi}^{(L,R)}\right) - \frac{1}{2}\right| \leq \frac{1}{2}\sqrt{1-4p(o^*)} \label{eq:app:ineq_lemma1}
\end{align}
\added{and if $p(o^*)=\frac{1}{4}$ then $p_X(o^*|\tilde\xi_1,\tilde\xi_2) >0$ if and only if $\widetilde\xi_1 \in S_{\xi_1}^L$ and $\widetilde\xi_2 \in S_{\xi_2}^R$.}

\begin{proof}
We define two sets for each of the hidden variables:
\begin{align}
    S^R_\alpha &:= \{\widetilde\alpha \mid \exists \widetilde\gamma: p(b = o^*\mid\widetilde\alpha\widetilde\gamma) > 0 \}, \quad S^L_\alpha = S_\alpha\backslash S^R_\alpha,  \nonumber\\ % S^L_\alpha := \{\widetilde\alpha \mid \exists \widetilde\beta: p(c = o^*\mid\widetilde\alpha\widetilde\beta) > 0 \}, \nonumber \\
    S^R_\beta  &:= \{\widetilde\beta  \mid \exists \widetilde\alpha: p(c = o^*\mid\widetilde\alpha\widetilde\beta) > 0 \}, \quad S^L_\beta = S_\beta\backslash S^R_\beta, \label{eq: App A: Subsets} \\ % S^L_\beta  := \{\widetilde\beta  \mid \exists \widetilde\gamma: p(a = o^*\mid\widetilde\beta\widetilde\gamma) > 0 \}, \label{eq: App A: Subsets} \\
    S^R_\gamma &:= \{\widetilde\gamma \mid \exists \widetilde\beta: p(a = o^*\mid\widetilde\beta\widetilde\gamma) > 0 \}, \quad S^L_\gamma = S_\gamma\backslash S^R_\gamma.  \nonumber% S^L_\gamma := \{\widetilde\gamma \mid \exists \widetilde\alpha: p(b = o^*\mid\widetilde\alpha\widetilde\gamma) > 0 \}, \nonumber
\end{align}
% \(p(S^L_\xi) \equiv \widetilde\xi_1,\) \(p(S^R_\xi) \equiv \widetilde\xi_2\) and
The sets \(S_\alpha\), \(S_\beta\), and \(S_\gamma\) denote the respective ranges of the variables \(\widetilde{\alpha}\), \(\widetilde{\beta}\), and \(\widetilde{\gamma}\). Furthermore, note that for any \(\xi\), the probabilities satisfy the relation
\(p(S^L_\xi) + p(S^R_\xi) = 1\)\added{, where we use the notation $p\left(\tilde\xi \in S_{\xi}^{L}\right) \equiv p(S_\xi^L)$}. Due to equations \eqref{eq:app:lemma1_conditions}, the possible outcomes with nonzero probability in this setup are, up to permutations
\begin{equation}
    (a,b,c) =
    \begin{cases}
        (\chi,\chi,\chi) \\
        (o^*,\chi,\chi)
    \end{cases}
    \label{eq: App A: possible cg outputs}
\end{equation}

To show that an $o^*$ outcome can only happen if a party receives one variable from an $R$ and one from an $L$, it is sufficient to assume the contrary. Assume that $p(a=o^*|\widetilde\beta\in S_R^\beta, \widetilde\gamma \in S_R^\gamma)>0$. However, this $\widetilde\beta$ value will sometimes result in an outcome $c=o^*$, since $\widetilde\beta\in S_\beta^R$, leading to a paradox (see Fig.~\ref{fig:cube_app_lemma1}).

To derive the inequality \eqref{eq:app:ineq_lemma1}, we formulate two inequalities based on the LHV model illustrated in Fig. \ref{fig:cube_app_constraint2}. Specifically, we express:
\begin{equation}
    p(S^R_\alpha) p(S^L_\beta) \geq p(\chi, \chi, o^*, S^R_\alpha \times S^L_\beta \times S^R_\gamma) + p(\chi, \chi, o^*, S^R_\alpha \times S^L_\beta \times S^L_\gamma) \equiv p(c=o^*)
    \label{eq: App A: Conditions for deriv. ineq in Lemma 1 - 1}
\end{equation}
\begin{equation}
    (p(S^R_\alpha) - 1) p(S^R_\beta) \leq - \left( p(o^*, \chi, \chi, S^L_\alpha \times S^R_\beta \times S^L_\gamma) + p(\chi,o^*,\chi, S^L_\alpha \times S^R_\beta \times S^R_\gamma) \right)
    \label{eq: App A: Conditions for deriv. ineq in Lemma 1 - 2}
\end{equation}
Rearranging inequalities \eqref{eq: App A: Conditions for deriv. ineq in Lemma 1 - 1}, \eqref{eq: App A: Conditions for deriv. ineq in Lemma 1 - 2} and using that all parties have the same probability of $o^*$ lead to
\begin{align}
    p(S^L_\beta)^2 - p(S^L_\beta) + p(o^*) \leq 0, \nonumber \\
    \Leftrightarrow \left|p\left(S_{\beta}^{L}\right) - \frac{1}{2}\right| \leq \frac{1}{2}\sqrt{1-4p(o^*)}
\end{align}
This derivation works similarly for all \(p(S^{(L,R)}_\xi)\). 
\end{proof}

\subsection{Linear Program}

\begin{figure}[t]
    \centering
    \includegraphics[width=0.85\linewidth]{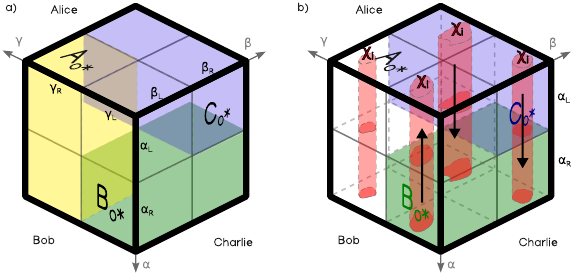}
    \caption{Cube representation of any LHV model which could possibly reproduce p(a,b,c). (a) The three cuboids $A_{o^*}, B_{o^*}, C_{o^*}$ are highlighted in yellow, green and blue, respectively. (b) For Constraint 3 we first notice that the $\chi_i$ responses on Alice's face define prisms (drawn approximately as cyclinders in this image for simplicity) within the cube, within which all the $a=\chi_i$ events must appear. We then consider the two prism parts which fall in $C_{o^*}$, and move down in both of them to reach $q(i,R)$ (roughly speaking) and a part in $B_{o^*}$. Then consider the two prisms parts in $B_{o^*}$ and move up in only one of them in order to reach $q(i,L)$. Substracting these quantities gives information on $q(i,L) - q(i,R)$ (up to some factors)while crucially canceling the prism part in the bottom right-most subcuboid.}
    \label{fig:cube_app_constraint2}
\end{figure}

In order to write constraints for a local hidden variable model, let us define certain regions of the local hidden variable model's cube representation, as well as variables which we will write the constraints for.
\begin{align}\label{eq:qdef}
    S_L:=& S_{\alpha}^L \times S_{\beta}^L \times S_{\gamma}^L,\\
    S_R:=& S_{\alpha}^R \times S_{\beta}^R \times S_{\gamma}^R, \\
    q(i,j,k,s):=& p(\chi_i, \chi_j,\chi_k, (\alpha,\beta,\gamma)\in S_s | (\alpha,\beta,\gamma)\in S_L \cup S_R),\\
    A_{o^*} :=& S_\alpha \times S_\beta^{L} \times S_\gamma^R,\\
    B_{o^*} :=& S_\alpha^R \times S_\beta \times S_\gamma^L,\\
    C_{o^*} :=& S_\alpha^L \times S_\beta^{R} \times S_\gamma,\\
    r(i,j,k,W):=& p(\chi_i, \chi_j,\chi_k, (\alpha,\beta,\gamma)\in W_{o^*}),
\end{align}
where $i,j,k\in\{1\dots|\chi|\}$, $s\in\{L,R\}$, $W\in\{A,B,C\}$. Note that the quantities $q(i,j,k,s)$ and $r(i,j,k,W)$ will be the variables in the linear program and the union of the (disjoint) sets $S_L, S_R, A_{o^*}, B_{o^*}, C_{o^*}$ gives the whole cube, as illustrated in Fig.~\ref{fig:cube_app_constraint2}(a). Moreover, to shorten notation we define $\alpha_L := |S_\alpha^L|$, $\alpha_R := |S_\alpha^R|$ such that $\alpha_L+\alpha_R=1$ and similarly for the others.

\textbf{Constraint 0} (normalization and positivity.)
\begin{align}
    q(i,j,k,s) &\geq 0 \;\;\;\forall i,j,k,s,\\
    \sum_{i,j,k,s} q(i,j,k,s) &= 1,\\
    r(i,j,k,W) &\geq 0 \;\;\;\forall i,j,k,W,
\end{align}

For the $r(i,j,k,W)$ variables the normalization is not 1\added{ as it is not defined as a conditional probability as the $q$ variables. Its normalization is related to the total amount of $\chi_i, \chi_j, \chi_k$ events, and thus to $q$, which we will now quantify. For this purpose, notice that t}\deleted{. T}he 
hidden variable groups $S_\xi^L$ and $S_\xi^R$ split the LHV model's cube representation in\added{to} 8 subcuboids, as shown in Fig. \ref{fig:cube_app_constraint2}. 
Notice that all $o^*$ events appear on the rectangles $S_{\xi_i}^L \times S_{\xi_{i+1}}^R$. We label the volumes behind these rectangles as $A_{o^*}, B_{o^*}, C_{o^*}$. Then the whole cube is divided into 5 regions: $A_{o^*}, B_{o^*}, C_{o^*}, S_L,S_R$. Thus
\begin{align}
    p(\chi_i, \chi_j,\chi_k) = p(\chi_i, \chi_j,\chi_k, A_{o^*}) + p(\chi_i, \chi_j,\chi_k, B_{o^*}) + p(\chi_i, \chi_j,\chi_k, C_{o^*}) + p(\chi_i, \chi_j,\chi_k, S_L) + p(\chi_i, \chi_j,\chi_k, S_R),
\end{align}
where we have used the shorthand $p(\chi_i, \chi_j,\chi_k, A_{o^*}) = p(\chi_i, \chi_j,\chi_k, (\alpha,\beta,\gamma)\in A_{o^*})$. By rearranging and using Eq.~(\ref{eq:qdef}), we arrive at\\
\textbf{Constraint 1.} (All $(\chi_i,\chi_j,\chi_k)$ outcomes are in one of the subcuboids.)
\begin{align}
    %\sum_s q(i,j,k,s) = \frac{1}{|S_L \cup S_R|}\big(p(\chi_i, \chi_j,\chi_k) - p(\chi_i, \chi_j,\chi_k, A_{o^*}) - p(\chi_i, \chi_j,\chi_k, B_{o^*}) - p(\chi_i, \chi_j,\chi_k, C_{o^*})\big).
    \sum_s q(i,j,k,s) = \frac{1}{|S_L \cup S_R|}\big(p(\chi_i, \chi_j,\chi_k) - \sum_{W} r(i,j,k,W)\big).
\end{align}
Notice that from Lemma 1, we know that e.g. in $A_{o^*}$ there can only be $(o^*,\chi,\chi)$ or $(\chi,\chi,\chi)$ events and $(o^*,\chi,\chi)$ events do not appear elsewhere. Hence $\beta_L \gamma_R = p(o^*,\chi,\chi) + p(\chi,\chi,\chi,A_{o^*})$. Thus we can establish that\\
\textbf{Constraint 2.} (marginals of $r$)
\begin{align}
    \sum_{i,j,k} r(i,j,k,A) &= \beta_L \gamma_R - p(a=o^*),\\
    \sum_{i,j,k} r(i,j,k,B) &= \gamma_L \alpha_R - p(b=o^*),\\
    \sum_{i,j,k} r(i,j,k,C) &= \alpha_L \beta_R - p(c=o^*).
\end{align}
Together with the inequality in Lemma 1, this implies that as $p(o^*)\rightarrow 1/4$, the variables $r(i,j,k,W)$ all go to 0, significantly simplifying the previous and following constraints, and recovering the structure of the LP in ~\cite{abiuso_single-photon_2022}. However, here we continue with the generic case. %\martin{r=0 also if p(o*) is not 1/4, e.g. for tilted QS in noiseless and single photon loss distribution, right?}

Finally, we wish to consider the marginals of $q(i,j,k,s)$, such as $q(i,s):=\sum_{j,k}q(i,j,k,s)$. For this purpose, we will write the occurrence of $a=\chi_i$ in both $B_{o^*}$ and $C_{o^*}$ in two ways so that $q$ appears in one of the forms. Then by subtracting the two we will get a constraint on $q(i,s=L)-q(i,s=R)$. We start by observing that since $b=o^*$ can only occur in $B_{o^*}$,
\begin{align}
    p(a=\chi_i,B_{o^*}) = p(a=\chi_i,b=o^*,B_{o^*}) &+ p(a=\chi_i,b\in\chi,B_{o^*}) = \\
    = p(a=\chi_i,b=o^*) &+ p(a=\chi_i,b\in\chi,c\in\chi,B_{o^*}).
\end{align}
The second way to write the occurrence of $\chi_i$ in $B_{o^*}$ is
\begin{equation}
    p(a=\chi_i,B_{o^*}) = p(a=\chi_i,B_{o^*}^{(L)}) + p(a=\chi_i,B_{o^*}^{(R)}) = \frac{\alpha_R}{\alpha_L}p(a=\chi_i,S_L) + p(a=\chi_i,B_{o^*}^{(R)}),
\end{equation}
where $B_{o^*}^{(L)} = \alpha_R \times \beta_L \times \gamma_L$ and $B_{o^*}^{(R)} = \alpha_R \times \beta_R \times \gamma_L$ are the two subcuboids of $B_{o^*}$ visibile in Fig.~\ref{fig:cube_app_constraint2}. In the last step we used that Alice's output can not depend on the hidden variable $\alpha$, hence, we can move from $B_{o^*}^{(L)}$ to $S_L$, provided we account for a factor $\frac{\alpha_R}{\alpha_L}$. One can write this with conditional probabilities, or simply by looking at the cube in Fig.~\ref{fig:cube_app_constraint2}, and realizing that the volume of the prism (approximately a cylinder in the drawing) which goes through both $B_{o^*}$ and $C_{o^*}$. The volume ratios of the top and bottom part of this prism is $\frac{\alpha_L}{\alpha_R}$.
Notice that the last term is related to the marginal of $q$ as $q(i,L) = \frac{1}{|S_L \cup S_R|}  p(a=\chi_i,S_L)$.

Similarly we can write these for $C_{o^*}$ as well.
\begin{align}
    p(a=\chi_i,C_{o^*}) = p(a=\chi_i,c=o^*) &+ p(a=\chi_i,b\in\chi,c\in\chi,C_{o^*}).
\end{align}
In the second equation we will not only use Alice's lack of knowledge of $\alpha$ for one term, but for both terms.
\begin{equation}
    p(a=\chi_i,C_{o^*}) = p(a=\chi_i,C_{o^*}^{(L)}) + p(a=\chi_i,C_{o^*}^{(R)}) = \frac{\alpha_L}{\alpha_R}p(a=\chi_i,B_{o^*}^{(R)}) + \frac{\alpha_L}{\alpha_R}p(a=\chi_i,S_R).
\end{equation}
Putting these together (eliminating $p(a=\chi_i,B_{o^*})$ and $p(a=\chi_i,C_{o^*})$) we have
\begin{align}
    \frac{\alpha_R}{\alpha_L}p(a=\chi_i,S_L) + \underline{p(a=\chi_i,B_{o^*}^{(R)})} = p(a=\chi_i,b=o^*) &+ p(a=\chi_i,b\in\chi,c\in\chi,B_{o^*})\\
    \frac{\alpha_L}{\alpha_R}\underline{p(a=\chi_i,B_{o^*}^{(R)})} + \frac{\alpha_L}{\alpha_R}p(a=\chi_i,S_R) = p(a=\chi_i,c=o^*) &+ p(a=\chi_i,b\in\chi,c\in\chi,C_{o^*}),
\end{align}
where we underlined the part that we have little information about and which we want to eliminate. We achieve this by multiplying the first equation by $\frac{\alpha_L}{\alpha_R}$ and then subtracting the second from the first to arrive at
\begin{align}
    p(a=\chi_i,S_L) - \frac{\alpha_L}{\alpha_R}p(a=\chi_i,S_R) =
    \frac{\alpha_L}{\alpha_R}p(a=\chi_i,b=o^*) - p(a=\chi_i,c=o^*) +\\  \frac{\alpha_L}{\alpha_R}p(a=\chi_i,b\in\chi,c\in\chi,B_{o^*}) - p(a=\chi_i,b\in\chi,c\in\chi,C_{o^*}) 
\end{align}
Finally, using the definitions $q(i,s) = \frac{1}{|S_L \cup S_R|}  p(a=\chi_i,S_s)$, $r(i,W) = p(\chi_i,\chi,\chi,W)$ and simplifying the notation we arrive at\\
\textbf{Constraint 3.} (marginals of $q$)
\begin{align}
    q(i,L) &= \frac{\alpha_L}{\alpha_R} q(i,R) + \frac{1}{|S_L \cup S_R|} \left( \frac{\alpha_L}{\alpha_R}p(\chi_i,o^*,\chi) - p(\chi_i,\chi,o^*) +  \frac{\alpha_L}{\alpha_R}r(i,B) - r(i,C)  \right)\\
    q(j,L) &= \frac{\beta_L}{\beta_R} q(j,R) + \frac{1}{|S_L \cup S_R|} \left( \frac{\beta_L}{\beta_R}p(\chi,\chi_j,o^*) - p(o^*,\chi_j,\chi) +  \frac{\beta_L}{\beta_R}r(j,C) - r(j,A)  \right)\\
    q(k,L) &= \frac{\gamma_L}{\gamma_R} q(k,R) + \frac{1}{|S_L \cup S_R|} \left( \frac{\gamma_L}{\gamma_R} p(o^*,\chi,\chi_k) - p(\chi,o^*,\chi_k) + \frac{\gamma_L}{\gamma_R} r(k,A) - r(k,B) \right).
\end{align}

The only thing remaining in order to turn these variables and constraints into a Linear Program is to determine $\alpha_L,\beta_L,\gamma_L$ (which determine the $\alpha_R, \beta_R,\gamma_R$ counterparts as well as the factor $1/|S_L \cup S_R|$). The most straightforward way is to apply the bounds from Lemma 1. The results provided in the maintext are obtained using this method. However, one can also use the grid technique, detailed in \cite{boreiri_noise-robust_2023} which gives even stronger constraints. The basic idea is that if the LP in infeasible for any triple $(\alpha_L,\beta_L,\gamma_L)$ then the original distribution must be nonlocal. However, testing all possible values of these variables is impossible technically, as they are continuous. The grid technique consists in dividing these continuous ranges into say \added{$M$}\deleted{$N$} segments, generating a 3D grid, where one can run the LP for each subcube in the grid by bounding $\alpha_L,\beta_L,\gamma_L$ to be within that subcube. By increasing \added{$M$}\deleted{$N$} one can approximate the ideal solution arbitrarily well. We have not implemented this technique (yet).

Note that using coarse grained distributions to reduce the size of the LP may be beneficial. We write about certain coarse graining in the next section.

\section{C. Distribution, coarse grainings and nonlocal regions}\label{app:distribution}
% \subsection{Output distribution for tilted quantum states}
\begin{figure}[t]
    \centering
    \includegraphics[width=0.30\linewidth]{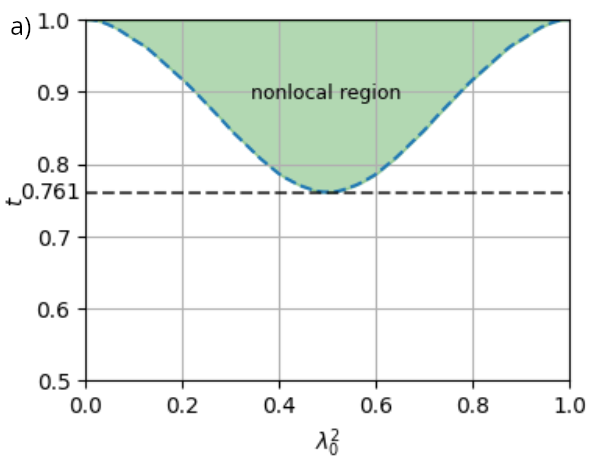}
    \includegraphics[width=0.32\linewidth]{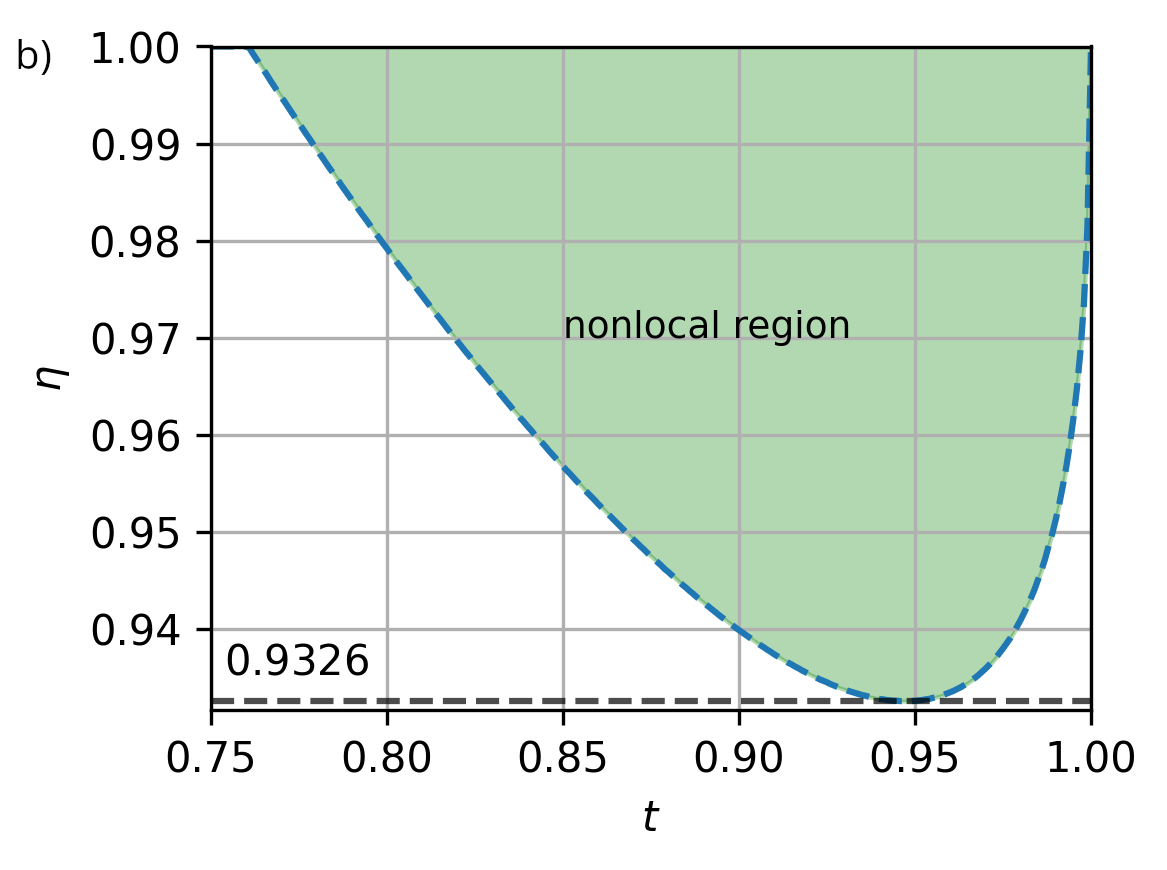}
    \includegraphics[width=0.32\linewidth]{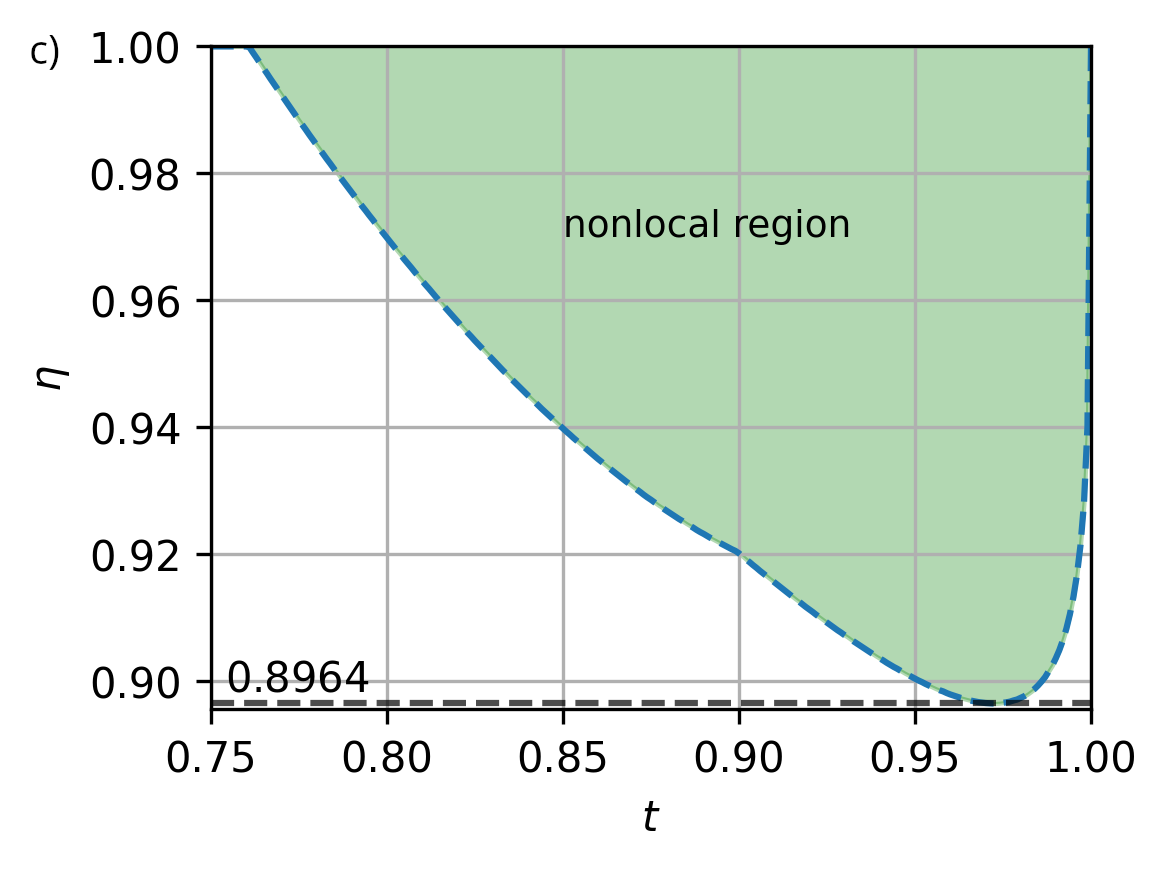}
    \caption{(a) Threshold \( t \) values beyond which the distribution \( p(a, b, c) \) does not admit a local model for a given \( \lambda_0^2 \). (b-c) Certified nonlocality of the (b) click/no-click and (c) photon number resolving distribution (events of 3 photons and higher coarse-grained) for the single photon loss model. For all plots $\varphi=\frac{\pi}{2}.$}
    \label{fig: Plot nonlocality range for N=2 photons}
\end{figure}

We outline the method used to compute the noiseless output distribution \( p(a, b, c) \) for NOON states, where the measurement outcomes are given by \( a, b, c \in \{0, N, 2N\} \) (or \( a, b, c \in \{0, L, R, 2\} \) in the case of click/no-click detectors). The transformation induced by a beamsplitter with transmissivity \( t \) and phase \( \varphi \) is expressed in terms of the creation operators \( a^\dagger \) as follows:
\begin{equation}
    \begin{pmatrix}
        a_2^\dagger \\
        a_1^\dagger
    \end{pmatrix}_i = 
    \begin{pmatrix}
        \sqrt{t} & -e^{-i\varphi} \sqrt{1-t} \\
        e^{i\varphi} \sqrt{1-t} & \sqrt{t}
    \end{pmatrix}_B 
    \begin{pmatrix}
        a_2^\dagger \\
        a_1^\dagger
    \end{pmatrix}_o.
\end{equation}
Using this transformation, we express the state transformation as %\tamas{maybe change the order in the kets?}
\begin{equation}
    |mn\rangle = \frac{1}{\sqrt{m!n!}} \left[ \sqrt{t} a_2^\dagger - e^{-i\varphi} \sqrt{1-t} a_1^\dagger \right]^m 
    \left[ e^{i\varphi} \sqrt{1-t} a_2^\dagger + \sqrt{t} a_1^\dagger \right]^n |00\rangle.
\end{equation}
For practical convenience, we define $u := \sqrt{t}$, $\added{l} := e^{-i\varphi} \sqrt{1-t}$ \deleted{ and $l^* := e^{i\varphi} \sqrt{1-t}$,} and rewrite the equation using the binomial theorem:
\begin{equation}
    |mn\rangle_{\text{in}} = \sum_{k=0}^{m} \sum_{q=0}^{n} \frac{m!}{k!(m-k)!} \frac{n!}{q!(n-q)!} \sqrt{(m+n-(k+q))!} \frac{\sqrt{(k+q)!}}{\sqrt{m!n!}} 
    (-1)^{m-k} \, u^{k+n-q} \, l^{m-k} \, l^{*q} |k + q, m + n - (k+q)\rangle_{\text{out}}.
    \label{eq: App: output states}
\end{equation}
The output state \( |.\rangle_{\text{out}} \) represents the photon number distribution at the output ports. To obtain the relevant POVMs $\{M^x_X\}$ for \( N=2 \) photons per source, we derive equations \eqref{eq: App: POVMs N=2 noiseless} $-$ \eqref{eq: App: POVMs N=2 noiseless - 22i}. % We set \(\tau := (1 - t)\).
\begin{gather}\label{eq: App: POVMs N=2 noiseless}
    |00\rangle_{\text{in}} = |00\rangle_{\text{out}}, \\[5pt]
    % |01\rangle_{\text{in}} = \sqrt{t} |01\rangle_{\text{out}} + \sqrt{\tau} e^{i\Phi} |10\rangle_{\text{out}}, \quad |10\rangle_{\text{in}} = -\sqrt{\tau} e^{-i\Phi} |01\rangle_{\text{out}} + \sqrt{t} |10\rangle_{\text{out}}, \nonumber \\[5pt]
    % |11\rangle_{\text{in}} = -\sqrt{2} \tau e^{-i\Phi} |02\rangle_{\text{out}} + (2t - 1) |11\rangle_{\text{out}} + \sqrt{2} t e^{i\Phi} |20\rangle_{\text{out}}, \nonumber \\[5pt]
    |02\rangle_{\text{in}} = t |02\rangle_{\text{out}} + \sqrt{2 t (1 - t)} e^{i\varphi} |11\rangle_{\text{out}} + (1 - t) e^{2i\varphi} |20\rangle_{\text{out}} \equiv v_1 |02\rangle_{\text{out}} + v_0 |11\rangle_{\text{out}} + u_1 |20\rangle_{\text{out}}, \label{eq: App: POVMs N=2 noiseless - 02i} \\[5pt]
    |20\rangle_{\text{in}} = (1 - t) |02\rangle_{\text{out}} - \sqrt{2 t (1 - t)} e^{-i\varphi} |11\rangle_{\text{out}} + t e^{-2i\varphi} |20\rangle_{\text{out}} \equiv v_{-1} |02\rangle_{\text{out}} - u_0 |11\rangle_{\text{out}} + u_{-1} |20\rangle_{\text{out}}, \label{eq: App: POVMs N=2 noiseless - 20i} \\[5pt]
    % |12\rangle_{\text{in}} = -t \sqrt{3} \tau e^{-i\Phi} |03\rangle_{\text{out}} + \left( t^2 - 2 \sqrt{t} \tau \right) |12\rangle_{\text{out}} + \left( 2t \sqrt{\tau} - \tau t^{\frac{3}{2}} \right) e^{i\Phi} |21\rangle_{\text{out}} + \sqrt{3} \tau t e^{2i\Phi} |30\rangle_{\text{out}},\nonumber \\[5pt]
    % |21\rangle_{\text{in}} = \sqrt{3} t \tau e^{-2i\Phi} |03\rangle_{\text{out}} - \left( 2t \sqrt{\tau} - \tau t^{\frac{3}{2}} \right) e^{-i\Phi} |12\rangle_{\text{out}} + \left( t^{3} - 2 \sqrt{t} \tau \right) |21\rangle_{\text{out}} + t \sqrt{3} \tau e^{i\Phi} |30\rangle_{\text{out}}, \nonumber \\[5pt]
    |22\rangle_{\text{in}} = \sqrt{6} t (1 - t) \left( e^{-2i\varphi} |04\rangle_{\text{out}} + e^{2i\varphi} |40\rangle_{\text{out}} \right) + \sqrt{6} \left( \sqrt{t} (1 - t)^{\frac{3}{2}} - t^{\frac{3}{2}} \sqrt{(1 - t)} \right) e^{-i\varphi} |13\rangle_{\text{out}} \label{eq: App: POVMs N=2 noiseless - 22i} \\ + \left( t^2 - 4t (1 - t) + (1 - t)^2 \right) |22\rangle_{\text{out}} + \sqrt{6} \left( t^{\frac{3}{2}} \sqrt{(1 - t)} - \sqrt{t} (1 - t)^{\frac{3}{2}} \right) e^{i\varphi} |31\rangle_{\text{out}}. \nonumber
\end{gather}
% Those equations allow us to construct the needed POVMs $\{M^x_X\}$.
From equations \eqref{eq: App: POVMs N=2 noiseless - 02i} and \eqref{eq: App: POVMs N=2 noiseless - 20i}, we can express the coefficients \(v_i\) and \(u_i\) in terms of the parameters \(t\) and \(\varphi\). To establish a clear relation between them, we first ensure that \(u_i\) depend only on the phase factor \(\varphi\). This is achieved by appropriately multiplying the coefficients \(u_0\) and \(v_0\) by a phase factor to maintain consistency in their definitions. With this adjustment, we define \(v_1 = t\), \(v_0 = \sqrt{2t(1-t)}\), and \(v_{-1} = (1 - t)\). Likewise, the coefficients \(u_i\) are given by \(u_1 = (1 - t) e^{2i\varphi}\), \(u_0 = \sqrt{2t(1-t)} e^{-2i\varphi}\), and \(u_{-1} = t e^{-2i\varphi}\). These expressions confirm that the relation
\begin{equation}
    v_i = |u_{-i}|, \quad i \in \{-1,0,1\}
\end{equation}
holds.
% This yields the following relation between the parameter \(u\) and \(v\)
% \begin{equation}
  %  v_i = |u_{-i}|, \,\, i\in \{-1,0,1\}.
% \end{equation}
As initial state we write tilted NOON-state as
\begin{equation}
    \ket{\psi}^{N=2}_{\xi} = \lambda_0\ket{02} + \lambda_1\ket{20},
    \label{eq: App: tilt. QS}
\end{equation}
with \(\lambda_0^2+\lambda_1^2=1 \,\, \text{and} \,\, \lambda_0,\lambda_1 \in\mathbb{R}\). The global initial state is given by
\begin{equation}
|\Psi\rangle_{A_1 A_2 B_1 B_2 C_1 C_2}^{N=2} \equiv |\psi\rangle^{N=2}_{A_2 B_1} \otimes |\psi\rangle^{N=2}_{B_2 C_1} \otimes |\psi\rangle^{N=2}_{C_2 A_1} = \lambda^3_0|202020\rangle + \lambda^2_0\lambda_1|002022\rangle + ... + \lambda^3_1|020202\rangle .
\label{eq: App C: initial state}
\end{equation}
% The distribution $p(a,b,c)$ is given by
% \begin{equation}
  %  p(a,b,c) = Tr\left( \ket{\Psi}^{N}\bra{\Psi}^{N} \cdot \left( 
% M^a_{A_1,A_2} \otimes M^b_{B_1,B_2} \otimes M^c_{C_1,C_2} \right) \right).
  %  \label{eq: App: p(a,b,c) = Tr(rho*M)}
% \end{equation}
Note that we abbreviate \(|202020\rangle\), which explicitly represents \(|20\rangle_{\text{in}} \otimes |20\rangle_{\text{in}} \otimes |20\rangle_{\text{in}}\). Now, we apply a coarse-graining of the outcomes to the set \(\{0, (2,0), (1,1), (0,2), 4\}\), meaning that we disregard the exact distribution of photons between modes when four photons are detected by a single party, i.e. \(|22\rangle_{\text{in}} = |22\rangle_{\text{out}}\). By substituting equations \eqref{eq: App: POVMs N=2 noiseless - 02i} and \eqref{eq: App: POVMs N=2 noiseless - 20i} into \eqref{eq: App C: initial state}, we obtain the global output state, leading to the following observations:
\begin{align}
    p(2_i,4,0) = \lambda_0^4\lambda_1^2 |u_i|^2,\;\; p(2_i,0,4) = \lambda_0^2\lambda_1^4 |v_i|^2, \label{eq: App C: p(2i,4,0) coarse grain output}\\
    p(2_i,2_j,2_k) = |\lambda_0^3 u_i u_j u_k + \lambda_1^3 v_i v_j v_k|^2.
    \label{eq: App C: p(2_i,2_j,2_k) coarse grain output}
\end{align}
Here, the outcomes \( 2_{-1}, 2_0, 2_1 \) correspond to the cases where the photon pair is distributed as \( (2,0) \), \( (1,1) \), or \( (0,2) \), respectively.

Note that the coarse graining can be broken up easily, e.g.
$p(2_i,(j,4-j),0) = p(2_i,4,0) \cdot \binom{4}{j} t^j (1-t)^{(4-j)}$. From the full statistics one can then derive the statistics for different coarse graining, for example the click/no-click distribution which corresponds to a physical setup, by coarse graining events as \{$(0,0),(i,0),(0,j),(i,j)$\} for $i>0,j>0$.
Other coarse grainings can also be interesting, in particular to reduce the number of outputs and thus the computational burden of the LP and LHV-Net. For example, for the noise robustness results with the LP we used the coarse graining \{0, (1,0), (0,1), (2,0), (1,1), (0,2), 4\}. Note that when we tried to reduce the number of 2-photon outcomes, e.g. by merging (0,2) and (1,1) or merging (0,2) and (2,0) outcomes, the LP always found a solution, i.e. it could not certify nonlocality. Finally, with LHV-Net besides examining the click/no-click distributions we also found nonlocality for a coarse graining of \{(2,0), (1,1), (0,2), Rest\}. This may be related to the PTC techniques introduced in~\cite{minimal_NL_boreiri_PhysRevA.107.062413}.

Based on Eqs. \eqref{eq: App C: p(2i,4,0) coarse grain output} and \eqref{eq: App C: p(2_i,2_j,2_k) coarse grain output}, we formulate a linear program (LP) to determine the nonlocal region as a function of the transmissivity \( t \) and the parameter \( \lambda_0^2 \), with the phase fixed at \( \varphi = \pi/2 \).

\section{D. Source Imperfection} \label{app:source_imperfections}

To set the stage for proving Lemma 2, we begin by introducing additional context regarding the operation of the sources. Specifically, rather than solely considering the distribution of photonic states at each source \(\xi\), we also incorporate the possibility of source failures. To model this, we associate to each source a failure bit \(f_\xi \in \{0,1\}\), where \(f_\xi = 0\) indicates that the source successfully generates a state, and \(f_\xi = 1\) denotes failure.\\
\\
\subsection{Proof of Lemma 2}

\added{\noindent\textbf{Lemma 2 (restated)} If each source $\xi$ can certify its failure, then by sending the additional failure bit $f_{\xi}$ to the parties they are connected to, the probability distribution $p\left((a,f_{\beta}, f_{\gamma}), (b,f_{\gamma}, f_{\alpha}), (c, f_{\alpha}, f_{\beta})\right)$ is local if and only if $p(a,b,c)$ is local.}

\begin{proof}
We coarse-grain the outcomes associated with the failure bits of the three sources \(\alpha, \beta, \gamma\) as \((F_{f_{\beta}+f_{\gamma}}, F_{f_{\alpha}+f_{\gamma}}, F_{f_{\alpha}+f_{\beta}})\), such that each party only announces the sum of the two incoming failure bits. The only possible outcomes, up to permutations, are:
\textbf{}
\begin{equation}
(F_{f_{\beta}+f_{\gamma}}, F_{f_{\alpha}+f_{\gamma}}, F_{f_{\alpha}+f_{\beta}}) = 
    \begin{cases}
        (F_0,F_0,F_0), \quad \text{all sources succeeded} \\
        (F_1,F_1,F_0), \quad \text{two sources succeeded} \\
        (F_2,F_1,F_1), \quad \text{one source succeeded} \\
        (F_2,F_2,F_2), \quad \text{no source succeeded}. \\
    \end{cases}
    \label{eq: App F: possible source failure outcomes}
\end{equation}
\begin{figure}[t]
    \centering
    \includegraphics[width=\linewidth]{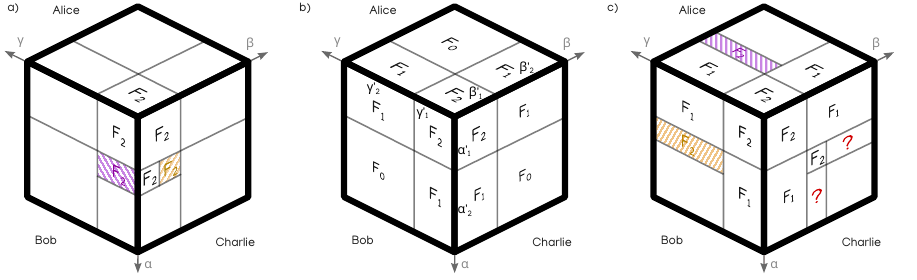}
    \caption{Cube representation of any LHV model which could possibly reproduce $p'\left((a_1,F_{f_{s_\beta} + f_{s_\gamma}}), (b_1,F_{f_{s_\gamma} + f_{s_\alpha}}), (c_1, F_{f_{s_\alpha} + f_{s_\beta}})\right)$. (a) Notice that for any maximal $(F_2,F_2,F_2)$ cuboid, the neighboring surfaces can not include any $F_2$ event. E.g. even if Charlie would have a tiny pixel of $F_2$ next to the $(F_2,F_2,F_2)$ cuboid, then the area marked by purple on Bob's side would also have to be $F_2$, which in turn implies that the yellow area on Charlie's face would also have to be $F_2$, thereby increasing the size of the ``maximal'' cuboid, arriving at a contradiction. (b) The only possible LHV structure with heralded source failures. Note that next to $F_2$'s there can not be $F_0$ events. (c) In the remaining parts, diagonal to the $F_2$ rectangles of the maximal $(F_2,F_2,F_2)$ cuboid there can only be $F_0$, If this were not the case, i.e. if there would be some $F_2$ on Charlie's side (note that $F_1$ is immediately ruled out since $(F_1,F_1,F_1)$ events are prohibited), then it would imply that $F_2$ appears on both Alice and Bob's sides (purple, yellow, resp.). This in turn creates a contradiction at the areas marked with ``\color{red}{?}\color{black}{'', where none of the events can be placed without contradiction.}}
    \label{fig:cube_app_lemma2}
\end{figure}

In the case of a local model considering only idealized, perfectly functioning sources, the model becomes trivial, as the only positive contribution corresponds to \((F_0,F_0,F_0)\), i.e., the local distribution \(p(a,b,c)\) is of the form (1) in the maintext.

For the more general case, we can establish a strict structure of local model in the cube picture. We start by rearranging the local hidden variables in order to form a cuboid of only $(F_2,F_2,F_2)$ events which is maximal, in the sense that no more complete blocks of $(F_2,F_2,F_2)$ appear behind it. More formally, let $([0,\widetilde\alpha],[0,\widetilde\beta],[0,\widetilde\gamma]$) be \deleted{the }hidden variables such that the responses of A, B and C are $(F_2,F_2,F_2)$. This cuboid is maximal if there does not exist an $\widetilde\alpha' > \widetilde\alpha$ such that $(\widetilde\alpha', \deleted{\beta,\gamma}\added{[0,\widetilde\beta],[0,\widetilde\gamma]})$ induces only $(F_2,F_2,F_2)$ events, and similarly for the other hidden variables. In terms of creating a maximal $(F_2,F_2,F_2)$ cuboid, one can start with any $(F_2,F_2,F_2)$ event and iteratively place other $(F_2,F_2,F_2)$ next to it by rearranging the hidden variables until one reaches a maximal one. At \deleted{one}\added{some} point one stops, as if there would exist an $\widetilde\alpha'$ as stated above then we could again just rearrange the $\alpha$ axis in order to create a slightly larger maximal $(F_2,F_2,F_2)$ cuboid.

In Fig.~\ref{fig:cube_app_lemma2}(a), we illustrate how starting from a maximal $(F_2,F_2,F_2)$ cuboid, one can easily see that beside it there can be no other $F_2$ events on any of the faces. If there were, then the cuboid would not be maximal. Hence all these parts must be either $F_1$ or $F_0$, however, as $(F_2,F_0,\cdot )$-type events are prohibited, it can only be $F_1$ immediately next to a maximal $(F_2,F_2,F_2)$, as shown in Fig.~\ref{fig:cube_app_lemma2}(b). Finally, in the remaining part, there can only be $F_0$ events, as we quickly run into a contradiction if there would be an $F_1$ (as $(F_1,F_1,F_1)$ is prohibited), and also if we put an $F_2$, as illustrated in Fig.~\ref{fig:cube_app_lemma2}(c). Hence, the only possible structure of the the cube for the \deleted{$a_2,b_2,c_2$}\added{failure bit} outcomes is given if Fig.~\ref{fig:cube_app_lemma2}(b) (as long as $p(F_0)>0$). The only question that remains is whether there is a local model including the \deleted{$a_1,b_1,c_1$}\added{other, original} outcomes for such a cube\added{, which essentially reduces to proving that the $(F_0,F_0,F_0)$ subcuboid can be scaled up to a full cube, and vice-versa}.

% In Fig. \ref{fig:cube_app_lemma2}, we present a LHV model in cube representation, which reproduces all possible outcomes listed in Eq. \eqref{eq: App F: possible source failure outcomes}.

A probability distribution \added{$p'$ }including the photonic variables \(a_1, b_1, c_1\) as well as the failure bits has a local model if it can be written as
\begin{multline}
    p'\left((a_1,F_{f_{\beta} + f_{\gamma}}), (b_1,F_{f_{\gamma} + f_{\alpha}}), (c_1, F_{f_{\alpha} + f_{\beta}})\right) = \\
    \int d\widetilde{\alpha} d\widetilde{\beta} d\widetilde{\gamma} \,\, p(\widetilde{\alpha},\widetilde{\beta},\widetilde{\gamma}) \,\, p(a_1,F_{f_{\beta} + f_{\gamma}}|\widetilde{\beta},\widetilde{\gamma}) \,\, p(b_1,F_{f_{\gamma} + f_{\alpha}}|\widetilde{\gamma},\widetilde{\alpha}) \,\, p(c_1,F_{f_{\alpha} + f_{\beta}}|\widetilde{\alpha},\widetilde{\beta}).
\end{multline}
For simplicity, let us refer to the failure bits, e.g., \( F_{f_{s_\beta} + f_{s_\gamma}} \), simply as \( a_2 \). By definition of the failure bits, it must hold that 
\begin{equation}
    p'\left( a_1 = a, b_1=b, c_1=c | a_2 = F_0, b_2 = F_0, c_2 = F_0 \right) = p(a,b,c).
    \label{eq: App F: relation betw. p' and p}
\end{equation}
Let us consider \( (F_0,F_0,F_0) \) to be within the subcuboid  
\begin{align}  
    [0,h_\alpha]\times[0,h_\beta]\times[0,h_\gamma],  
    \label{eq: App F: F_0 subcube}  
\end{align}  
where \( h_\xi = 1-p(f_\xi) \) represents the probability of successful heralding of a source. Furthermore, the variables \( \widetilde\alpha, \widetilde\beta, \widetilde\gamma \) are uniformly distributed random variables in the interval \([0,1]\). Note that compared to Fig. \ref{fig:cube_app_lemma2}, here we have inverted the hidden variables \( \widetilde{\xi} \to 1-\widetilde{\xi} \). \\
(\( \impliedby \)):\\
We assume that \(p(a,b,c)\) is local. Then there exists a local model
\begin{align}
    p(a,b,c) = \int d\widetilde\alpha d\widetilde\beta d\widetilde\gamma & p(a|\widetilde\beta,\widetilde\gamma)  p (b|\widetilde\gamma, \widetilde\alpha) p(c|\widetilde\alpha,\widetilde\beta).
\end{align}
We define
\begin{equation}
    p'(x_1,x_2|\widetilde{\xi_1},\widetilde{\xi_2}) := 
    \begin{cases}
        p(x_1|\frac{\widetilde{\xi_1}}{h_{\xi_1}}, \frac{\widetilde{\xi_2}}{h_{\xi_2}}) \cdot \delta_{x_2, F_0}, & \text{if} \,\, \widetilde{\xi_1} \leq h_{\xi_1}, \,\, \widetilde{\xi_2} \leq h_{\xi_2} \\
        \delta_{x_1,0} \cdot \delta_{x_2, F_1}, & \text{if} \,\, \widetilde{\xi_1} > h_{\xi_1}, \,\, \widetilde{\xi_2} \leq h_{\xi_2} \,\, \text{or} \,\, \widetilde{\xi_1} \leq h_{\xi_1}, \,\, \widetilde{\xi_2} > h_{\xi_2} \\
        \delta_{x_1,0} \cdot \delta_{x_2, F_2}, & \text{if} \,\, \widetilde{\xi_1} > h_{\xi_1}, \,\, \widetilde{\xi_2} > h_{\xi_2}
    \end{cases}
    \label{eq: App F: Proof Lemma 2 - cases}
    \end{equation}

Using this defined response function, we can construct a local model. However, we must be sure that using \eqref{eq: App F: Proof Lemma 2 - cases} we \deleted{write }satisfy the constraint (\ref{eq: App F: relation betw. p' and p}).
\begin{equation}
    p'(a_1=a,a_2=F_0,b_1=b,b_2=F_0,c_1=c,c_2=F_0) = \int_0^1\int_0^1\int_0^1 d\widetilde{\alpha}d\widetilde{\beta} d\widetilde{\gamma} \,\, p(\widetilde{\alpha},\widetilde{\beta},\widetilde{\gamma}) \,\, p'(a,F_0|\widetilde{\beta} \widetilde{\gamma})p'(b,F_0|\widetilde{\alpha} \widetilde{\gamma})p'(c,F_0|\widetilde{\beta} \widetilde{\alpha}) = \nonumber
\end{equation}
\begin{equation}
=\int_0^{h_\alpha}\int_0^{h_\beta}\int_0^{h_\gamma}d\widetilde{\alpha}d\widetilde{\beta} d\widetilde{\gamma} \,\, p\left(\frac{\widetilde{\alpha}}{h_\alpha},\frac{\widetilde{\beta}}{h_\beta},\frac{\widetilde{\gamma}}{h_\gamma}\right) p'\left(a\mid 
 \frac{\widetilde{\beta}}{h_\beta},\frac{\widetilde{\gamma}}{h_\gamma}\right)p'\left(b\mid\frac{\widetilde{\alpha}}{h_\alpha},\frac{\widetilde{\gamma}}{h_\gamma}\right)p'\left(c\mid \frac{\widetilde{\alpha}}{h_\alpha},\frac{\widetilde{\beta}}{h_\beta}\right) =
  \label{eq: App F: Proof lemma 2 <==}
\end{equation}
\begin{equation}
    = h_\alpha h_\beta h_\gamma \int_0^1\int_0^1\int_0^1 d\widetilde{\alpha'}d\widetilde{\beta'} d\widetilde{\gamma'} \,\, p(\widetilde{\alpha'},\widetilde{\beta'},\widetilde{\gamma'}) \,\, p'(a|\widetilde{\beta'},\widetilde{\gamma'}) p'(b|\widetilde{\alpha'},\widetilde{\gamma'}) p'(c|\widetilde{\alpha'},\widetilde{\beta'}) = p'(a_2=F_0,b_2=F_0,c_2=F_0)  \cdot p(a,b,c) \nonumber
\end{equation}
In \eqref{eq: App F: Proof lemma 2 <==}, the first step comes from realizing that the $F_0$'s imply an upper bound on the variables \( \widetilde{\alpha}, \widetilde{\beta}, \widetilde{\gamma} \), namely by \( h_\alpha, h_\beta, h_\gamma \), respectively. In the second step, we performed a change of variables, setting \( \xi' := \frac{\widetilde{\xi}}{h_\xi} \).
\\
The reverse direction (\( \implies \)) works analogously but one must use the response functions of \(p'\) to define those of \(p\) as
\begin{equation}
    p(x_1|\widetilde\xi_1,\widetilde\xi_2):= p'(x_1=x,x_2=F_0|h_{\xi_1}\widetilde{\xi_1},h_{\xi_2}\widetilde{\xi_2}).
\end{equation}
Then
\begin{align}
    &p(a,b,c) = \int_0^1\int_0^1\int_0^1 d\widetilde{\alpha}d\widetilde{\beta} d\widetilde{\gamma} \,\, p'(a_1=a,a_2 = F_0|\widetilde{\beta} h_\beta,\widetilde{\gamma} h_\gamma)
    p'(b_1=b,b_2 = F_0|\widetilde{\gamma} h_\gamma,\widetilde{\alpha} h_\alpha)
    p'(c_1=c,c_2 = F_0|\widetilde{\alpha} h_\alpha,\widetilde{\beta} h_\beta) =\\
    &\int_0^{h_\alpha''}\int_0^{h_\beta''}\int_0^{h_\gamma''} d\widetilde{\alpha''}d\widetilde{\beta''} d\widetilde{\gamma''} (h_\alpha h_\beta h_\gamma)^{-1}\,\, p'(a_1=a,a_2 = F_0|\widetilde{\beta''},\widetilde{\gamma''})
    p'(b_1=b,b_2 = F_0|\widetilde{\gamma''},\widetilde{\alpha''})
    p'(c_1=c,c_2 = F_0|\widetilde{\alpha''},\widetilde{\beta''}) =\\
    &\int_0^{1}\int_0^{1}\int_0^{1} d\widetilde{\alpha''}d\widetilde{\beta''} d\widetilde{\gamma''} (h_\alpha h_\beta h_\gamma)^{-1}\,\, p'(a_1=a,a_2 = F_0|\widetilde{\beta''},\widetilde{\gamma''})
    p'(b_1=b,b_2 = F_0|\widetilde{\gamma''},\widetilde{\alpha''})
    p'(c_1=c,c_2 = F_0|\widetilde{\alpha''},\widetilde{\beta''}) =\\
    &p'(a_1=a,b_1=b,c_1=c|a_2=F_0,b_2=F_0,c_2=F_0),
\end{align}
where in the first equation we have used the newly defined response functions, in the second we have changed the integration variables, in the third we have used for example that $p(a_1=a,a_2=F_0|\beta'' > h_\beta \text{ or } \gamma''>h_\gamma )=0$, hence we can add the missing parts of the whole cube to the integral, as these sum up to 0.
\end{proof}

\subsection{Proposed Setup for State Generation}
\begin{figure}[t]
    \centering
        \includegraphics[height=0.24\textwidth]{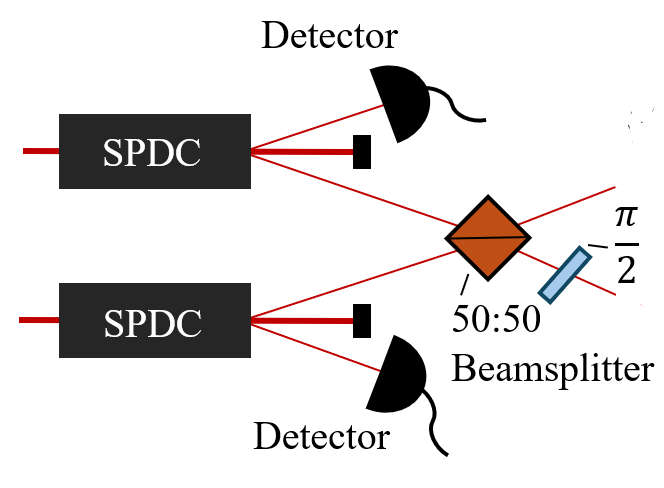}
        \includegraphics[height=0.24\textwidth]{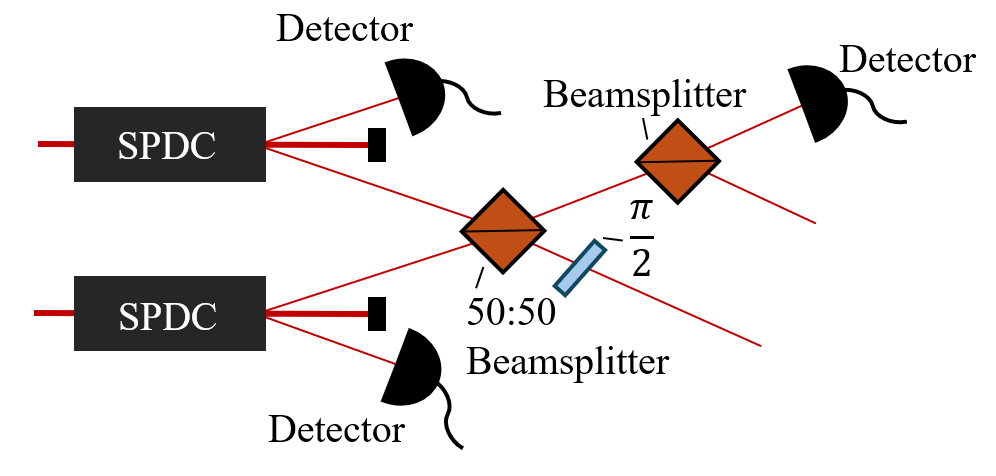}
    \caption{Proposed state generation setups: (Left) Heralded generation of the balanced entangled state using two SPDC sources, photon detectors, as well as a 50:50 beamsplitter and a phase shifter to produce the state \(\frac{1}{\sqrt{2}} (|02\rangle + |20\rangle)\). (Right) Modified setup for generating tilted quantum states. An additional beamsplitter is introduced in one output mode, and the third mode is measured to post-select the desired outcome, allowing control over the degree of tilt.}
    \label{fig: Proposed Setup for State Generation}
\end{figure}
We now proceed to apply Lemma 2 to propose a\deleted{n}\added{ proof of principle} implementation that uses heralded generation of entanglement. Specifically, we focus on utilizing heralded sources based on spontaneous parametric down-conversion (SPDC) to produce balanced quantum states while enforcing the certification of source failure events.
The setup consists of two independent SPDC sources, each designed to produce photon pairs. One photon from each pair is directed to a heralding detector, while the other photon propagates toward the parties the source is connected to. Successful detection of a heralding photon signifies the creation of spatially-entangled quantum state in the output mode. Upon simultaneous heralding from both SPDC sources, the remaining photons are interfered at a beam splitter. Due to the Hong–Ou–Mandel effect, this interference ideally results in the generation of the entangled state:
\[\frac{1}{\sqrt{2}} (|02\rangle + |20\rangle).\]
To integrate the failure certification mechanism outlined in Lemma 2, we introduce a classical bit signal associated with each SPDC source. Each source includes a shutter that blocks the photon path to the parties unless both heralding events are successful. If that is the case, the source opens the shutter and the source transmits a bit value '0' alongside the photon. Conversely, if either heralding fails, the source transmits a classical bit '1' instead of sending a photon. In this manner, the failure bit \(f_\xi\) explicitly certifies the success or failure of each source.
This configuration allows us to interpret the bit values as tokens in the token-counting framework. Note that, the source emits either two photons accompanied by '0' bits or zero photons accompanied by two '1' bits. As a result, the photon number distribution remains consistent with token-counting principles, and the same nonlocality arguments apply to the output distribution.

To extend the setup towards generating tilted quantum states, we introduce a controlled modification by placing an additional beam splitter in one of the output modes after the initial interference at the first beamsplitter. The extra beam splitter introduces an additional mode, transforming the state into \(\frac{1}{\sqrt{2}} (a |011\rangle + b |002\rangle + c |020\rangle) + \frac{1}{\sqrt{2}} |200\rangle\), where the coefficients $a,b,c$ depend on the transmissivity and reflectivity of the second beam splitter. The third mode is then measured, and events with one or two photons in this mode are discarded. This post-selection projects the system onto \(\frac{1}{\sqrt{2}} c |020\rangle + \frac{1}{\sqrt{2}} |200\rangle\). After discarding the trivial third mode and renormalizing, the resulting two-mode state is
\[\ket{\psi_{\text{tilt}}} = \frac{c}{\sqrt{c^2 + 1}} \ket{02} + \frac{1}{\sqrt{c^2 + 1}} \ket{20},\]
where the degree of tilt is tunable via the beamsplitter’s transmissivity, encapsulated in the parameter \(c\).

\added{Note that in a practical implementation of the scheme a central question could be the rate at which 6-fold single photon generation events occur. Following the back-of-the-envelope reasoning and the values used in Ref.~\cite{abiuso_single-photon_2022}, one would arrive at an experiment time on the order of magnitude of a year to achieve about 3000 events (a similar number of events as in Ref.~\cite{wang2024experimentalgenuinequantumnonlocality}) with the proof of principle scheme we proposed. Naturally, certain experimental setups could be operating with better parameters than what was used in Ref.~\cite{abiuso_single-photon_2022}, potentially making it already realizable. However, experimental design principle could also greatly reduce the time needed. For example one could use multiplexing to increase the rate of photons from a single source, or one could use different sources such as on-demand photon emitters based on quantum dots. If we would herald such sources, then rates may be on the same order of magnitudes. However, given that these sources have an exceptionally high probability of emitting a photon based on an input, it may make sense to treat all generation requests as quasi-heralded photon generations and incorporate the non-generation events into the photon loss model. Furthermore, one could combine different source generation techniques (heralding some of the sources while treating the non-generation of other sources as photon loss). Either way, given the rapid pace of technological improvement and the specialization available in laboratories, an experiment may very well be within reach. This, however, will be determined by the specific characteristics of the devices used.}

\section{E. LHV-Net implementation}\label{app:LHVNet}

\begin{figure}[t]
    \centering
        \includegraphics[width=\textwidth]{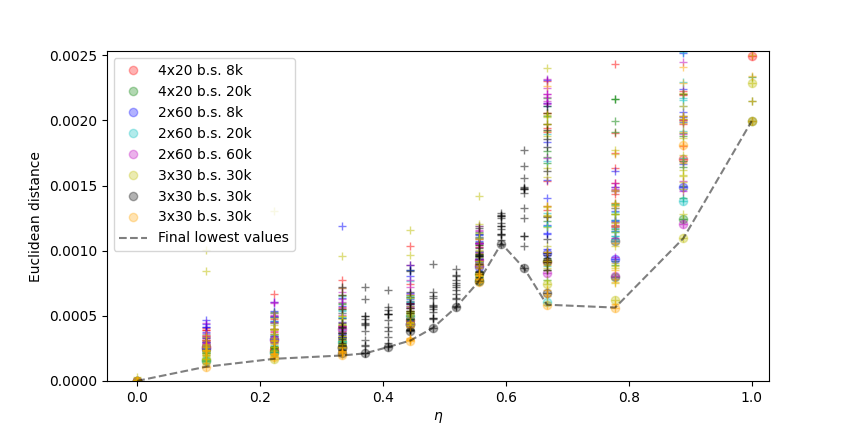}
    \caption{Results of 930 trainings of the neural network from scratch (denoted by ``+'')\added{ for $t=0.75, \varphi=\pi/2, Q=0$}, with different depth $\times$ width and batch sizes (b.s. in legend). The lowest value of a given training set is shown with a circle, while the minimal over all points is shown with a dashed line (and is what is portrayed in the maintext in Fig.~2).}
    \label{fig:AppFigNN}
\end{figure}

\added{One may rightfully ask whether the resurgent peak in the neural network scan in Fig.~2 in the maintext or in Fig.~\ref{fig:AppFigNN} here reflects the true physics or is just a shortfalling of the neural network. We first discuss how these results were obtained and then discuss why we believe them to be trustworthy.}

The neural networks used for finding local models consist of three smaller multilayer perceptrons, one for each party, each receiving their respective inputs (see~\cite{krivachy_neural_2020} for details). We took the base size of these multilayer perceptrons to be width 30-60, depth 2-4 (see Fig.~\ref{fig:AppFigNN} for the different configurations used). The number of samples of local hidden variables (the ``batch size'') was taken to be between 8000 and 30000. The neural networks were trained in epochs of 5000 batches, using an Adadelta optimizer with a Kullback-Leibler divergence as loss for the first epoch (the first epoch was done twice and the best one was kept). From the second epoch stochastic gradient descent was run on this loss, while from the 7th epoch, the L2 distance was used as a loss for fine-tuning the Euclidean distance. When switching to the L2 distance, Adadelta was used again for 2 epochs, after which stochastic gradient descent was used to fine-tune the model for another 16 epochs.

In order to avoid local minima, the above procedure was done multiple times and the best distance was kept. Moreover, some parameters such as width, depth and batch size were modified, particularly in an attempt to make the resurgent peak disappear. This, however, did not move. A summary of 930 trainings (each one a repetition of the above procedure for a given set of neural network parameters and $\eta$ value) can be seen in Fig.~\ref{fig:AppFigNN}.

\added{
There are several hints, which together give a strong indication that the observed peak is truly there and is not an artifact of the numerical neural-network method.}

\added{Indications of high noise robustness even without considering the neural network’s results:}
\begin{enumerate}
\item \added{The LP certifies robustness to about 10\% single-photon loss even though it only uses partial information: in particular it does not require anything specific regarding the internal structure of the local model inside the $S_L$ and $S_R$ parts of the cube. It is already known that in the single photon case the estimated true noise robustness is $\sim10$ times larger than what a similar LP can certify ($\sim5\%$ vs $\sim0.5\%$\cite{abiuso_single-photon_2022,boreiri_noise-robust_2023}). Also, we know that if we would use the extra information within the $S_L$ and $S_R$ cubes, it would allow to unlock nonlocality in new regimes (see Ref.~\cite{pozas_continuous_families}). In our case using this extra information is computationally quite expensive, as it can be integrated in the LP technique with the inflation method, which requires exceptionally large memory. Nonetheless, these inflation techniques are currently being developed and improved, hopefully soon they can easily be applied to the current scenario as well. All in all, based on these, even without considering the peak that we observed in Fig. 2, we expect that a much higher noise robustness can be certified than what our LP certifies.}
\item \added{High-dimensional entanglement has proved to be more noise robust in many other quantum communication scenarios, hence it is not so surprising that we get higher noise robustness.}
\item \added{When the $\ket{02} + \ket{20}$ state loses a photon, it closely resembles a dephased single photon source $\ket{01} + \ket{10}$. Hence, we expect that it would be relatively robust in the regime where single photon loss is dominant. Given that two-photon loss is approximately $(1-\eta)^2$ and single photon is approximately $2(1-\eta)\eta$, we expect a transition between the two regions to appear somewhere above $\eta \approx 1/3$.}
\end{enumerate}

\added{Indications of high noise robustness when considering the neural network’s results:}
\begin{enumerate}\setcounter{enumi}{3}
\item \added{The neural network has been tested many times on distributions with such output cardinalities (4x4x4), in which cases it has never contradicted our theoretical knowledge, and in fact has led to surprising conjectures which have since been proven. Hence, we believe the way we use the neural network (doing many independent runs and keeping the best results) has been reasonably benchmarked and can be used with a high degree of confidence (see e.g. [\cite{krivachy_neural_2020,abiuso_single-photon_2022,supic_genuine_NNL2022,pozas_full_network_nonlocality_2022,boreiri_krivachy_sekatski_topRobNL_2024,minimal_NL_boreiri_PhysRevA.107.062413}]).}
\item \added{In Fig. \ref{fig:AppFigNN} one can see the resulting Euclidean distances of many neural network runs, with different parameters for the neural networks. Notice that at the peak the distance is around 0.001. At slightly different $\eta$ values, the neural networks manage to consistently find lower values for the Euclidean distances than 0.001. In particular this means that for target distributions similar to the one at the peak it does find smaller distances consistently - for a variety of different neural network architectures. Hence, it is not the case that the neural network in general fails for this region of $\eta$ values, but it is really at this target distribution at the peak where there is no local distribution which is close.
Put otherwise: \textit{if} we would see that in this $\eta$ region the neural network would be having difficulties going beneath 0.001 for other target distributions as well, then we could say that we need to do more runs so the neural network can also go below 0.001 for the peak. This is, however, not the case.}
\end{enumerate}

\end{document}